\documentclass[prb,twocolumn,showpacs,preprintnumbers,amsmath,amssymb]{revtex4}
\usepackage{graphicx}% Include figure files
\usepackage{dcolumn}% Align table columns on decimal point
\usepackage{bm}% bold math

\def\k#1{| #1 \rangle}
\def\b#1{\langle #1 |}
\def\bk#1#2{\langle #1 | #2 \rangle}
\def\kb#1#2{| #1 \rangle\langle #2 |}
\def\band#1{\scriptscriptstyle(#1)}

\def\oiint{\int\@oiint\int}
\def\@oiint{%
   \mathchoice
      {\mkern-18mu\bigcirc\mkern-18mu}%
      {\mkern-12mu\circ\mkern-12mu}%
      {\mkern-12mu\circ\mkern-12mu}%
      {\mkern-12mu\circ\mkern-12mu}}%

\begin{document}

\preprint{TQCP-CondMatt}

\title{Topological quantization by controlled paths: application to Cooper pairs pumps}% Force line breaks with \\

\author{Raphael Leone}
\email{Raphael.Leone@grenoble.cnrs.fr}
\author{Laurent L\'evy}
\affiliation{ Institut N\'eel, C.N.R.S.- Universit\'e Joseph Fourier, BP 166,
38042 Grenoble-cedex 9, France}

\date{\today}% It is always \today, today,

\begin{abstract}
When physical systems are tunable by three classical parameters, level
degeneracies may occur at isolated points in parameter space.  A topological
singularity in the phase of the degenerate eigenvectors exists at these points.
When a path encloses such point, the accumulated geometrical phase is sensitive
to its presence.  Furthermore, surfaces in parameter space enclosing such point
can be used to characterize the eigenvector singularities through their Chern
indices, which are integers. They can be used to quantize a physical quantity
of interest. This quantity changes continuously during an adiabatic evolution
along a path in parameter space. Quantization requires to turn this path into a
surface with a well defined Chern index.  We analyze the conditions necessary
to a {\em Topological Quantization by Controlled Paths}. It is applied to
Cooper pair pumps. For more general problems, a set of four criteria are
proposed to check if topological quantization is possible.
\end{abstract}


\maketitle

\section{\label{sec:intro}Introduction}
Using nanometer size Josephson junctions, a huge variety of superconducting
quantum circuits can be made.  These circuits are described by simple
Hamiltonians involving a discrete set of quantum variables.  They are typically
the excess number of Cooper pairs ${\hat n}_j$ on superconducting elements and
their canonical conjugate variables $\hat{\Theta}_j$
($[\hat{n}_j,\hat{\Theta}_j]=i$) are related to the quantum phases of the
superconducting order parameters of the circuit islands. In addition, most
circuits have tunable elements: they are control voltages on gates, or using
magnetic fluxes, quenched quantum phases or Josephson couplings. These circuits
are most often used to implement quantum logic\cite{Nakamura99,Vion02}, where
the quantum gates are controlled with voltage or resonant microwave pulses on
some of gates or other tunable elements.

The tunable elements of quantum circuits can also be used to generate adiabatic
evolutions of the Hamiltonian as a function of the parameters. More precisely,
the $N$ induced gate charges $n_{gi}$ and control phases $\varphi_i$ define a
vector $\mathbf{R}=\{n_{g1}\ldots\,\varphi_1\ldots\}$ in the parameter space
$\mathbb{P}$ of dimension $N$.  Let $E_\alpha(\mathbf{R})\ldots$ and
$\k{\alpha(\mathbf{R})}\ldots$ be the eigenenergies (bands) and eigenvectors of
the Hamiltonian $\hat{\mathcal{H}}(\mathbf{R})$. At a point $\mathbf{R}$ and
for a non-degenerate band $\alpha$, one can construct a fiber defined by the
set of all vectors $\{\k{\alpha(\mathbf{R})}\}$ which may differ by a complex
factor. The set of all these fibers defines the fiber bundle over the parameter
space $\mathbb{P}$. When the topology of this bundle becomes non-trivial,
physical phenomena of great interest can occur.

Parallel transport, holonomy and homotopy are central concepts for the physics
of geometric phases. Berry's phase is one of these and is a relevant quantity
when adiabaticity conditions holds for a globally non-degenerate band. In this
case, Berry's phase is the geometric part of the phase acquired by the
wavefunction along an adiabatic cycle over a closed path $\Gamma^c$  in the
parameter space $\mathbb{P}$.

Depending on the nature of the quantum system studied, the physical
consequences of the non-trivial topology of the eigenvector bundle are
different. A number of physical examples have been studied in several areas of
physics. In molecular systems, the electronic structure depends on the
semi-classical nuclear coordinates (within the Born-Oppenheimer approximation)
which define the parameter space. Their energy manifolds can have conical
intersections at isolated values of the nuclear
coordinates~\cite{Herzberg63,Faure00,Zhilinskii01}: these so-called
``diabolical points'' are directly responsible for the change of multiplicity
of rotation-vibrations levels as a function of nuclear coordinates. In
molecular magnets, the magnetic energy levels depend on the direction and
magnitude of the applied magnetic field (the parameter space) with respect to
the molecular axes.  For some molecules, isolated degeneracies have also been
found for specific direction and values of the magnetic
field~\cite{Wernsdorfer99,Bruno06}. At these points, quantum tunnelling is
quenched as a result of interferences caused by the wavefunction phase changes
around the ``defect''. Indeed, this phase change takes a particular value of
$\pi$ for physically relevant paths encircling the degeneracy, leading to the
destructive interferences observed.

In the few examples above, degeneracies occur at isolated conical
intersections between two energy bands in a three-dimensional
parameter space. These diabolical points are singularities of the
quantum phase field over the parameter space and responsible for
the ``exotic topologies''. Closed paths, through Berry's phase,
are sensitive probes of the topology. Closed surfaces in a
three-dimensional parameter space are also sensitive to the
presence of conical points through a topological invariant called
the Chern index $c_1$, which is an integer number. Some of the
best known phenomena in condensed matter physics are
well-understood in terms of Chern indices, such as integer quantum
Hall effect (IQHE)~\cite{Thouless82,Kohmoto84}, Thouless Pumping
(TP)~\cite{Thouless83} or AC Josephson effect
(ACJE)~\cite{Thouless83,Goryo07}. Physical quantities which can be
expressed in term of Chern indices are subject to
\textit{topological quantization}. This is why they are used in
metrology. For instance, the IQHE gives a conversion from voltage
unit (Volt) to current unit (Amp\`ere) through the resistance
quantum $R_K=\frac{h}{e^2}$: $V=\frac{R_H}{c_1}I$ ; in the same
way, the AC Josephson effect gives a conversion from voltage to
frequency (in Hertz) through the magnetic flux quantum
$\Phi_0=\frac{h}{2e}$: $V=\Phi_0\,\nu$.

Such physics can be encountered in the simplest superconducting
circuit depending on three \textit{tunable} parameters: the Cooper
Pair Pump (CPP), where degeneracies occur at isolated points in
the parameter space. Here, the parameter space is constructed from
two gate voltages $V_{gi}$ and a quenched quantum phase $\varphi$.
However, an essential difference subsists between the examples
given above and our problem. In the case of IQHE, TP or ACJE, the
relevant physical quantity measured is directly proportional to
the Chern index of the surface brought into play (the magnetic
Brillouin zone for the IQHE). In a CPP, one can only make
(adiabatic) paths in the parameter space by modifying the
parameters in order to tune the current delivered by the CPP.
Thus, Berry's phase seems to be, \textit{a priori}, the relevant
topological quantity characterizing the paths. This will be shown
to always be the case, by relating the charge transferred by the
CPP through a path is always expressible in term of a Berry's
phase, even for open paths. Nevertheless, Chern indices can also
specify the value of the current for specific paths covering
densely a closed surface enclosing a diabolical point. The degree
of quantization of the current delivered by this method improves
exponentially with the degree of the surface coverage. This form
of quantization is referred to as ``Topological Quantization by
Controlled Paths'' (TQCP). The current will be shown to be equal
to $2e\,\nu$ where $\nu$ is a characteristic frequency of the
adiabatic cycles of pumping. For metrology, the Cooper pair
pumping through TQCP gives a conversion between current and
frequency: it is an effect which could be used to close the
metrological triangle between the units of voltage $V$, current
$I$ and frequency $\nu$.

In this paper, we emphasize what is new and specific to quantum
circuits, and the example of the Cooper pair pump is an excellent case
study for the concept of TQCP around which much of the paper is built.
In Sec. \ref{sec:fiber-bundle}, we recall the
 topological properties of three dimensional parameter spaces
stressing the notions of Berry's phase and Chern indices in the
presence of degeneracies. In Sec. \ref{sec:TQCP}, the TQCP is
introduced, and the computation method of the quantized physical
quantity is given. In this section, the necessary criteria for
TQCP are derived, namely
\begin{enumerate}
\item The energy spectrum must be discrete (adiabaticity).
\item The Hamiltonian depends on three continuous \textit{tunable}
parameters, which specify the parameter space $\mathbb{P}$. Isolated
\textit{conical degeneracies} between the two lowest eigenstates $\k{\pm}$ must
occur in $\mathbb{P}$.
\item The relevant physical observable $Q$ (the quantity
measured) follows Hamilton semiclassical equation of motion ${\dot
Q}=\langle\partial_{\varphi}\hat{\mathcal{H}}\rangle$ where the parameter
$\varphi$ is periodic. The contribution of this quantity along {\em
geometrical} paths is set by the topology of the eigenvector bundle.
\item The dynamical contributions to $Q$ must also
be taken into account. The topological quantization can be implemented only
when they can be eliminated. This is possible when the $\varphi$-dependence of
the Hamiltonian eigenvalues can be integrated out using its periodicity or
other symmetries of the system.
\end{enumerate}
In the conclusion, a full discussion of these four criteria is presented in
light of this work. Section \ref{sec:charge-quantization} is devoted to a
practical implementation of the TQCP for Cooper pair pumps. Section
\ref{sec:adiab-computing} shows how microwave fields can be used to expand the
parameter space to higher dimensions. In the example considered, the isolated
degeneracies become a two dimensional degenerate subspace in which non-abelian
holonomies are designed for adiabatic quantum computation.

\section{\label{sec:fiber-bundle}Topology of the parameterized Eigenvectors Space}

In this section, the topological features in parameter spaces are explained in
simple words. Let a quantum system be dependent of $N$ parameters $x^\mu$
defining a parameter space $\mathbb{P}$. Then, the Hamiltonian governing the
dynamics is written as $\hat{\mathcal{H}}(\mathbf{R})$, where
$\mathbf{R}=(x^1,x^2,\dots,x^N)$ is a vector in $\mathbb{P}$. The parameters
are classical and can be tuned by an observer. Modifying the parameters amounts
to trace a path $\Gamma$ in the parameter space, parameterized by time. To each
point $\mathbf{R}$ in $\mathbb{P}$ is assigned the set
$\{\k{\beta(\mathbf{R})}\}$ of eigenstates of $\hat{\mathcal{H}}(\mathbf{R})$,
with the dimensionality of the Hilbert space $\mathcal{E}$. More precisely, for
a single non-degenerate level $\alpha$, a \textit{fiber}
$\mathcal{F}^{\band{\alpha}}_{\mathbf{R}}$ attach the eigenvector
$\k{\alpha(\mathbf{R})}$ to the point $\mathbf{R}$ as:
\begin{align}
\mathcal{F}^{\band{\alpha}}_{\mathbf{R}}=\Big\{\k{\alpha}\;\textrm{such~that}\;
\hat{\mathcal{H}}(\mathbf{R})\k{\alpha}=E_\alpha(\mathbf{R})\k{\alpha}\Big\}\,.
\label{bundle}
\end{align}

%For instance, if
%$\k{\alpha(\mathbf{R})}\in\mathcal{F}^{\band{\alpha}}_{\mathbf{R}}$,
%then for all complex number $\lambda$:
%$\lambda\k{\alpha(\mathbf{R})}\in\mathcal{F}^{\band{\alpha}}_{\mathbf{R}}$.

This fiber is defined everywhere in $\mathbb{P}$, except where the band
$\alpha$ is degenerate. The set of all fibers attached to $\mathbb{P}$ defines
the \textit{vector bundle} $\mathcal{F}^{\band{\alpha}}$ over the parameter
space: for a non-degenerate band $\alpha$, it is a complex line bundle. A
\textit{connection} is a differentiable rule for a shift from the fiber
$\mathcal{F}^{\band{\alpha}}_{\mathbf{R}}$ to
$\mathcal{F}^{\band{\alpha}}_{\mathbf{R}+d\mathbf{R}}$ when $\mathbf{R}$ moves
to $\mathbf{R}+d\mathbf{R}$ in parameter space. When considering the adiabatic
evolution of a quantum state $\k{\tilde\alpha(\mathbf{R})}$, \textit{parallel
transport connections} are involved.  They are such that
${\Im}m\,\bk{\tilde{\alpha}(\mathbf{R})}{d\tilde{\alpha}(\mathbf{R})}=0$
everywhere along the path $\Gamma$ covered.  This requires that the path
$\Gamma$ never crosses a point where the band $\alpha$ is degenerate. For a
smooth choice of normalized states $\k{\alpha(\mathbf{R})}$, the parallel
transport condition on a state
$\k{\tilde{\alpha}}=e^{i\gamma_\alpha}\k{\alpha}$ is equivalent to a time
evolution of the phase $\dot{\gamma}_\alpha=i\bk{\alpha}{\dot{\alpha}}$ which
can be integrated along the path $\Gamma$ starting from $\mathbf{R}_i$ as
\begin{align}
\gamma_\alpha(t)=i\int_{\mathbf{R_i}}^{\mathbf{R(t)}}
\bk{\alpha(\mathbf{R})}{\boldsymbol{\nabla}\alpha(\mathbf{R})}\cdot
d\mathbf{R}\,.
\label{phasegamma}
\end{align}
This phase is purely geometric, i.e. independent of a
reparameterization of coordinates on the path $\Gamma$.  On the
other hand, it is not invariant under a local gauge change:
$\k{\alpha}\to e^{i\xi(\mathbf{R})}\k{\alpha}$, making the phase
$\gamma_\alpha$ non integrable and multi-valued.  For this reason,
the gauge field and its Berry's connection are specified as
\begin{align}
\mathbf{A}^{\band{\alpha}}(\mathbf{R})&=
\bk{\alpha(\mathbf{R})}{\boldsymbol{\nabla}\alpha(\mathbf{R})},\\
\mathcal{A}^{\band{\alpha}}&=\bk{\alpha}{d\alpha}=
\mathbf{A}^{\band{\alpha}}(\mathbf{R})\cdot d\mathbf{R}.
\label{vecpot}
\end{align}
A key feature of the phase $\gamma_\alpha$ is that it becomes a
gauge invariant quantity when the paths $\Gamma^c$ are closed. In
this case, \textit{Berry's phase}~\cite{Berry84}
\begin{align}
\gamma_\alpha(\Gamma^c)=\oint_{\Gamma^c}\mathcal{A}^{\band{\alpha}}\mod
[2\pi]
\end{align}
is a physically observable quantity and cannot be removed by any
local gauge change.  It is sensitive to the topology of the fibre
bundle:  it is the \textit{holonomy}~\cite{Simon83} of the line
bundle $\mathcal{F}^{\band{\alpha}}$ over the path $\Gamma^c$. An
important special case arises when the Hamiltonian is real for a
set of paths in a subspace of $\mathbb{P}$: a continuous choice of
real eigenstates $\k{\tilde{\alpha}(\mathbf{R})}$ may be chosen
over this path, which defines a parallel transport since
${\Im}m\,\bk{\tilde{\alpha}}{d\tilde{\alpha}}=0$, leading to
values of $0$ or $\pi$ for Berry's phase, such that
$\k{\tilde{\alpha}(\mathbf{R_f})}=\pm\k{\tilde{\alpha}(\mathbf{R_i})}$.

Berry's phase first appeared as the geometric contribution to the
phase acquired in the adiabatic cyclic evolution of a
non-degenerate state in the parameter space. In the next section,
the state $\k{\psi_\alpha(t)}$ of a system initially prepared in
the non-degenerate state $\k{\alpha(\mathbf{R_i})}$ evolving
adiabatically along the path $\Gamma$ is shown to be
approximatively
\begin{align}
\k{\psi_\alpha(t)}\approx
e^{-i\eta_\alpha(t)}e^{i\gamma_\alpha(t)}\k{\alpha(\mathbf{R}(t))}
\label{adiabaticastate}\,,
\end{align}
where $\eta_\alpha(t)=\frac{1}{\hbar}\int_0^t E_\alpha(t') dt'$ is the usual
dynamical phase and $\k{\alpha(\mathbf{R})}$, the instantaneous eigenstate.

The gauge field $\mathbf{A}^{\band{\alpha}}$ defined in
Eq.~\ref{vecpot} is analogous to the vector potential of
electromagnetism: in three dimensions, the gauge insensitive
magnetic field
$\mathbf{B}^{\band{\alpha}}=\boldsymbol{\nabla}\times\mathbf{A}^{\band{\alpha}}$
is physically relevant.  To characterize the properties of the
fiber bundle in a gauge independent manner, it is useful to define
Berry's curvature as the differential form
$\mathcal{B}^{\band{\alpha}}=B^{\band{\alpha}}_{\mu\nu}\,dx^\mu
dx^\nu\equiv i\b{d\alpha}\wedge\k{d\alpha}$, where
$B_{\mu\nu}^{\band{\alpha}}=\partial_\mu
A^{\band{\alpha}}_\nu-\partial_\nu A^{\band{\alpha}}_\mu$ are
elements of the antisymmetric curvature tensor
$B^{\band{\alpha}}$. Using these definitions, Stokes theorem can
be used to write the Berry's phase of the band $\alpha$ over the
closed path $\Gamma^c$ as a surface integral
\begin{align}
\gamma_\alpha(\Gamma^c)=\oint_{\Gamma^c}\mathcal{A}^{\band{\alpha}}=
\iint_{\mathcal{S}}\mathcal{B}^{\band{\alpha}}\,,
\label{flux-formula}
\end{align}
where $\mathcal{S}$ is an oriented surface with $\Gamma^c$ as a border. Systems
depending on a set of 3 parameters $x^1,x^2,x^3$ are most pertinent for TQCP.
The orientation of $\mathbb{P}$ is defined by a local choice of basis, for
instance the natural basis $(\mathbf{u}_1,\mathbf{u}_2,\mathbf{u}_3)$ with
respect to the coordinates $x^\mu$. Using ordinary vector calculus, the
antisymmetric curvature tensor $B^{\band{\alpha}}$ reduces to a magnetic field
$\mathbf{B}^{\band{\alpha}}$, the curl of $\mathbf{A}^{\band{\alpha}}$.
$\mathbf{B}^{\band{\alpha}}$ can be computed directly from the Hamiltonian
gradient as\cite{Berry84}
\begin{align}
\label{magnetique2}
\mathbf{B}^{\band{\alpha}}=i\sum_{\beta\ne\alpha}
\frac{\b{\alpha}\boldsymbol{\nabla}\hat{\mathcal{H}}\k{\beta}\times\b{\beta}\boldsymbol{\nabla}\hat{\mathcal{H}}\k{\alpha}}{\big(E_\alpha-E_\beta\big)^2}\,,
\end{align}
where $\{\k{\beta}\}$ is the set of eigenstates of $\hat{\mathcal{H}}$,
dependant on $\mathbf{R}$. As we shall see below, nonzero Berry's
phases occur from a non-trivial topology of the eigenvectors fiber
bundle.   This occurs at level degeneracies in the parameter space
where the magnetic field $\mathbf{B}^{\band{\alpha}}$ is singular. At
these points
$\boldsymbol{\nabla}\cdot\mathbf{B}^{\band{\alpha}}\ne\mathbf{0}$ and
$\mathbf{A}^{\band{\alpha}}$ cannot be defined. Without such points,
the topology is trivial and the parallel transport leaves states
invariant over a closed loop:
$\k{\tilde{\alpha}(\mathbf{R_f})}=\k{\tilde{\alpha}(\mathbf{R_i})}$.

The von Neumann-Wigner theorem\cite{vonNeumann29} asserts that in a 3D
parameter space, accidental degeneracies may occur between two neighboring
levels (say $\k{\pm}$) only at isolated points $\mathbf{R_i}^*$; these
degeneracies which are singularities of the fields $\mathbf{B}^{\band{\pm}}$
have been named \textit{normal singular points} by Simon\cite{Simon83}. Since
the gauge fields $\mathbf{A}^{\band{\pm}}$ do not exist at these points, they
become local quantities. Interesting physics appear when such points live in
$\mathbb{P}$. It can be visualized most easily by projecting the Hamiltonian on
the two-level manifold $\k{\pm}$ which becomes degenerate at the singular point
$\mathbf{R}^*=(x^1_*,x^2_*,x^3_*)$.  Using the projector
$\hat{P}=\kb{+}{+}+\kb{-}{-}$ a gradient expansion of the two-level projection
$\hat{\mathcal{H}}_{\pm}$ of the Hamiltonian can be made in the vicinity of the
singularity $\mathbf{R}^*$
\begin{align}
\hat{P}\big[\hat{\mathcal{H}}(\mathbf{R})-\hat{\mathcal{H}}(\mathbf{R}^*)\big]\hat{P}&=
\hat{\mathcal{H}}_\pm(\mathbf{R})-\hat{\mathcal{H}}_\pm(\mathbf{R}^*)\notag\\
&=\boldsymbol{\nabla}\hat{\mathcal{H}}_{\pm}(\mathbf{R}^*)\cdot\delta\mathbf{\mathbf{R}}+O(\delta\mathbf{R}^2),
\end{align}
where $\delta\mathbf{R}=\mathbf{R}-\mathbf{R^*}=(\delta x^1,\delta x^2,\delta
x^3)$. With a suitable choice for the origin of energies
($E_\pm(\mathbf{R}^*)=0$), $\hat{\mathcal{H}}_\pm(\mathbf{R}^*)$ is zero.  With
this choice, this expansion can be expressed on the basis of Pauli matrices as
\begin{align}
\hat{\mathcal{H}}_\pm(\mathbf{R})=\frac{1}{2}\,\sum_{\mu,\nu=1}^{3}c_\mu^\nu\delta
x^\mu\sigma_\nu\,.
\end{align}
The $c_\mu^\nu$ are the elements of a $3\times 3$ real matrix $\hat{C}$ which
has a nonzero determinant for linear level crossing at $\mathbf{R}^*$.  This
becomes more familiar by defining
$\mathbf{b}=\hat{C}\delta\mathbf{R}=(b^x,b^y,b^z)$ as the effective magnetic
field for an equivalent spin-$\frac{1}{2}$ spin system,
\begin{align}
\hat{\mathcal{H}}_\pm(\mathbf{R})=\frac{1}{2}\,\boldsymbol{\sigma}\cdot\mathbf{b}(\mathbf{R})=
\frac{1}{2}\left(\begin{array}{cc}b^z &
b^x-ib^y\\b^x+ib^y&-b^z\end{array}\right)
\label{Zeeman}
\end{align}
which magnitude increases linearly with the deviation from the degeneracy point
$\mathbf{b}^*=\mathbf{0}$.  The energy levels $E_\pm = \pm
\frac{|\mathbf{b}|}{2}$ of the two bands intersect conically at the degeneracy
point (also called \textit{conical point} or \textit{diabolical point}). The
matrix $\hat C$ maps a local neighborhood of $\mathbf{R}^*$ of the parameter
space onto a spatially isotropic spin-$\frac{1}{2}$ hamiltonian in the magnetic
field $\mathbf{b}$. As long as the mapping amounts to a local deformation of
the parameter space, and no additional degeneracies appear in the vicinity of
$\mathbf{R}^*$, the topology of the fiber bundle stays unchanged. If $\hat{C}$
changes the orientation of space (the $\textrm{det}(\hat{C})$ is negative), the
sign of the topological charge is flipped by the mapping.  The one-to-one
mapping $\hat C$ allows to use the Euler angles of $\mathbf{b}$ rather than the
coordinate $\delta \mathbf{R}$ to specify the eigenstates in the vicinity of
the singularity as
\begin{align}
\label{spineurs}
\begin{array}{l}|+(\mathbf{b})\rangle=
\left(\begin{array}{c}\cos\frac{\theta}{2}\\e^{i\phi}\sin\frac{\theta}{2}\end{array}\right),
|-(\mathbf{\mathbf{b}})\rangle=
\left(\begin{array}{c}-\sin\frac{\theta}{2}\\e^{i\phi}\cos\frac{\theta}{2}\end{array}\right)\end{array}\!.
\end{align}
For both levels, one can assign Berry's gauge potentials
$\mathbf{A}^{\band{\pm}}$,
\begin{align}
\mathbf{A}^{\band{\pm}}=i\bk{\pm}{\boldsymbol{\nabla}_{\mathbf{b}}\pm}=
\pm\frac{\cos\theta\mp1}{2|\mathbf{b}|\sin\theta}\,\mathbf{e}_\phi\,.
\end{align}
These are the azimuthal gauge fields of ``Dirac monopoles'' of strength
$+\frac{1}{2}$ or $-\frac{1}{2}$ placed at the origin. They are singular on
their \textit{Dirac string} ($\theta=\pi$ for $\k{+}$ and $\theta=0$ for
$\k{-}$).  These monopoles produce a radial magnetic field of opposite
directions
\begin{align}
\mathbf{B}^{\band{\pm}}=\boldsymbol{\nabla}_{\mathbf{b}}\times\mathbf{A}^{\band{\pm}}=
\mp\frac{\mathbf{b}}{2|\mathbf{b}|^3}\label{magfield}\,.
\end{align}
The strengths $\pm\frac{1}{2}$ are more easily identified by
taking the divergence
\begin{align}
\boldsymbol{\nabla}_{\mathbf{b}}\cdot\mathbf{B}^{\band{\pm}}=
\mp\frac{1}{2}\,\delta(\mathbf{b})\,,
\end{align}
which integrated over any volume including the origin gives
$\pm\frac{1}{2}\times4\pi$. Since the one to one mapping ${\hat C}$ between
parameter space and spin-space conserves the flux, the topological charge in
$\mathbb{P}$ is preserved up to a sign (when ${\hat C}$ changes the surfaces
orientations).

\begin{figure}
\includegraphics[width=0.8\columnwidth]{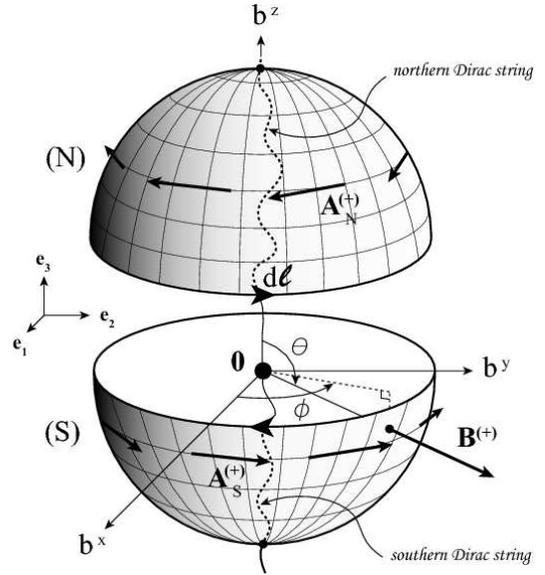}
\caption{A monopole placed at the degeneracy points generates azimuthal vector
potentials $\mathbf{A}^{\band{\pm}}$ for each quantum band $\k{\pm}$
intersecting at the monopole. Because of the singularity, a single valued
expression for $\mathbf{A}^{\band{\pm}}$ exists separately in the upper and
lower half hemispheres. They can be connected on the equator by a clutching
function ($\mathbf{A}^{\band{+}}$ is pictured here). The corresponding magnetic
fields $\mathbf{B}^{\band{\pm}}$ are radial and decrease as $b^{-2}$, where $b$
is the distance to the singularity.} \label{half-spheres}
\end{figure}

The degeneracies ($\mathbf{R_i}^*$) in parameter space appear as singularities
of the fields $\mathbf{A}^{\band{\pm}}$. Since any surface $\mathcal{S}$
enclosing $\mathbf{R}^*$ intersects the Dirac string, $\mathbf{A}^{\band{\pm}}$
is not defined everywhere on $\mathcal{S}$.  It is possible to make
$\mathbf{A}^{\band{\pm}}$ single valued only by making a hole in $\mathcal{S}$
through which the Dirac string can be threaded: in this case, the surface can
be continuously contracted to a point without crossing the singularity
$\mathbf{R}^*$.  This is the reason why there is no single analytic expression
of $\mathbf{A}^{\band{\pm}}$ over a surface which encloses completely the
degeneracy.  An alternative procedure for defining $\mathbf{A}^{\band{\pm}}$
was made by Wu and Yang\cite{Wu76}. The space is divided in north $(N)$ and
south $(S)$ halves (see Fig.~\ref{half-spheres}), with a different gauge choice
$\mathbf{A}^{\band{\pm}}$ in each part, which are related by an appropriate
\textit{clutching function} $f^{\band{\pm}}$ on the equator where the
eigenstates are connected using $\k{\pm}_N=e^{if^{\band{\pm}}}\k{\pm}_S$, with
$\mathbf{A}_N^{\band{\pm}}=\mathbf{A}_S^{\band{\pm}}+\boldsymbol{\nabla}\,f^{\band{\pm}}$.
For the isotropic spin-$\frac{1}{2}$ model, the different determination of
$\mathbf{A}^{\band{\pm}}$ are:
\begin{align}
\mathbf{A}^{\band{\pm}}=\pm\left\{\begin{array}{lll}\displaystyle\frac{\cos\theta-1}{2|\mathbf{b}|\sin\theta}=
\mathbf{A}_N^{\band{\pm}}&\textrm{for}\;\;\theta\in\Big[0\,;\displaystyle\frac{\pi}{2}\Big)\\
\\
\displaystyle\frac{\cos\theta+1}{2|\mathbf{Y}|\sin\theta}=
\mathbf{A}_S^{\band{\pm}}&\textrm{for}\;\;\theta\in\Big(\displaystyle\frac{\pi}{2}\,;\pi\Big]&
\end{array}\right.,
\end{align}
with $f^{\band{\pm}}=\mp\,\phi$ as clutching function.

\begin{figure}
\includegraphics[width=\columnwidth]{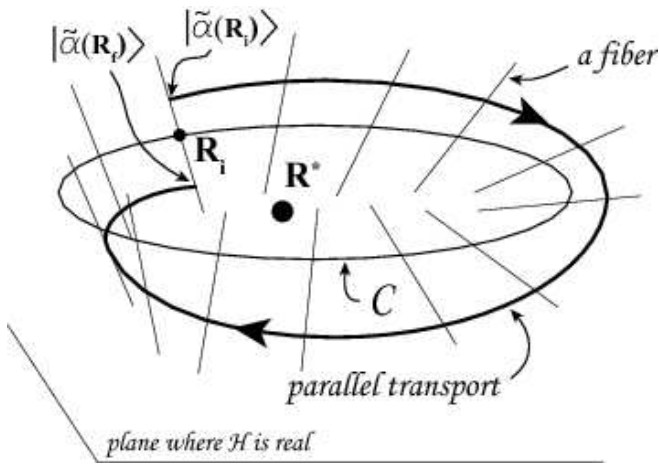}

\caption{Illustration of a circular path $\mathcal{C}$ around a degeneracy,
lying in a plane where the Hamiltonian is real. The real eigenstates depend on
a single angle variable.  This angle specifies the direction of a line. Under
parallel transport along loop, this line (the eigenvector) generates a
M\oe{}bius strip which is the visual representation of the eigenvector bundle.
The eigenstate changes sign when coming back to the same point. This sign is
the holonomy of the eigenstate fiber.} \label{mobiusstrip}
\end{figure}

On a closed path $\Gamma^c$, Berry's phase for the two-levels $\pm$ is
sensitive to the presence of a degeneracy at the origin since
\begin{align}
\gamma_\pm(\Gamma^c)=\oint_{\Gamma^c}\mathbf{A}^{\band{\pm}}\cdot
d\boldsymbol{\ell}=\mp\frac{1}{2}\,\Omega(\Gamma^c)\,,
\end{align}
where $\Omega(\Gamma^c)$ is the solid angle seen from the origin.
When $\Gamma^c$ is contained in a plane intersecting the origin
then $\gamma_\pm(\Gamma^c)$ is just equal to $\pi$ times the
winding number of $\Gamma^c$ around the origin. This discussion
makes it clear that it is a consequence of the nontrivial topology
of the bundles $\mathcal{F}^{\band{\pm}}$ around the origin.  A
geometrical illustration is possible when the path $\mathcal{C}$
shown in Fig. \ref{mobiusstrip} lies into a plane where the
Hamiltonian is real: the spin-eigenstates $\k{\pm}$, which can be
taken as real, depends on a single angle variable (say $\theta$)
which defines a line. As one moves along $\Gamma^c$ this line,
which represents the eigenvector bundle, covers a one-twist
M\oe{}bius strip as illustrated in Fig.~\ref{mobiusstrip}. In this
parallel transport, initial and final states are seen to be
opposite $\k{\pm(\mathbf{b_f)}}=-\k{\pm(\mathbf{b_i})}$: Berry's
phase equals to $\pi$. It is also the well-known property of the
group $SU(2)$, where rotations are $4\pi$ periodic. In general, if
the path encircles $m$ degeneracies and has a winding number $n_i$
around the degeneracy at the points $\mathbf{R}^*_i$, Berry's
phase is $0 \mod[2\pi]$ if $\sum_{i=1}^mn_i$ is even and $\pi$
otherwise.

Berry's phase factors for loops in planes where the Hamiltonian is
real are \textit{topological invariants} $\pm 1$, which
characterize the sum of winding numbers around degeneracies.
Another topological invariant is obtained after integrating the
field $\mathbf{B}^{\band{\pm}}$ of Eq.~\ref{magfield} over a small
sphere $\mathbb{S}^2$ around the origin. When normalized to
$2\pi$, these flux gives the first Chern numbers (or \textit{Chern
indices}) of the sphere with respect to the two bands $\k{\pm}$:
\begin{align}
c_1^{\band{\pm}}(\mathbb{S}^2)=\frac{1}{2\pi}\oiint_{S^2}\mathbf{B}^{\band{\pm}}\cdot
\hat{n}\,dS=\mp\, 1\,,\label{chernflux}
\end{align}
where $\hat{n}$ is the unit vector normal to the surface of the sphere. The
Chern index can also be computed from the potentials $\mathbf{A}^{\band{\pm}}$
using
\begin{align}
c_1^{\band{\pm}}(\mathbb{S}^2)&=\frac{1}{2\pi}\oint_{\mathcal{C}}\big(\mathbf{A}^{\band{\pm}}_N-\mathbf{A}^{\band{\pm}}_S\big)\cdot
d\boldsymbol{\ell}=\frac{1}{2\pi}\oint_{\mathcal{C}}\boldsymbol{\nabla}_{\mathbf{b}}f^{\band{\pm}}\cdot
d\boldsymbol{\ell}\notag\\
&=\mp\frac{1}{2\pi}\int_0^{2\pi}\!\!\!
d\phi=\mp\frac{1}{2\pi}\,(2\pi-0)=\mp\,1\,.
\end{align}
The mapping $\hat C$ between the parameter space $\mathbb{P}$ and the isotropic
space $\mathbb{R}^3$ does not change Chern indices if space orientation is
preserved ($\textrm{det}(\hat{C})>0$).

Topological indices do not depend on the projection on a two-level system,
which is valid only in a small neighborhood of $\mathbf{R}^*$. For any band
$\alpha$ and any closed surface $\mathcal{S}$ in $\mathbb{P}$, Gauss theorem
assures that the integral of $\mathbf{B}^{\band{\alpha}}$ over the entire
surface is identical to the sum of the integrals over a small sphere about each
degeneracy. In other words,
\begin{align}
c_1^{\band{\alpha}}(\mathcal{S})=\sum_{d_i\in
V(\mathcal{S})}\,q_i^{\band{\alpha}}\,,
\end{align}
where $q_i^{\band{\alpha}}$ represents the topological charge of
each degeneracy $\mathbf{R_i}^*$ inside the volume
$V(\mathcal{S})$. The Chern index does not depend on the geometry
of the closed surface and is a topological invariant which depends
only on the degeneracies it contains.

\section{\label{sec:TQCP}Topological Quantization by Controlled Paths}
Suppose that a quantum system depends on three \textit{tunable} parameters
$x^1$, $x^2$ and $x^3$ which specify the space $\mathbb{P}$. One can always
construct two angles
$\vartheta^1(\mathbf{R}),\vartheta^2(\mathbf{R})\in[0,2\pi)$ in $\mathbb{P}$
which parameterize a two-dimensional torus $\mathbb{T}^2$ and
$\boldsymbol{\vartheta}=(\vartheta^1,\vartheta^2)$ is a vector on
$\mathbb{T}^2$.  $Q$ is a physical quantity which can be expressed as the
partial derivative of the Hamiltonian
$\hat{\mathcal{H}}(\boldsymbol{\vartheta})$ with respect to one of the angles
(say $\vartheta^2$)
\begin{align}
\dot{Q}=\big\langle\partial_2\hat{\mathcal{H}}(\boldsymbol{\vartheta})\big\rangle=
\,\b{\psi}\partial_2\hat{\mathcal{H}}(\boldsymbol{\vartheta})\k{\psi}\,.
\label{formedebase}
\end{align}
TQCP can only be used for such physical observable, which is followed
adiabatically on a path $\Gamma_0$ lying on the torus.  Physically, it
is the ground state  expectation value of $Q$ which is of interest. On
the path $\Gamma_0$, there will be one or more avoided level crossings
with other levels, and Zener tunnelling in their vicinity sets the
rates of variation for the parameters required for adiabaticity.  Using
the spin representation (Eq.~\ref{Zeeman}) close to a level crossing,
where $b_z(t)$ is the tuning parameter, the condition for
adiabaticity\cite{Landau32} may be written as
\begin{align}
\hbar\, \dot{b}_z(t) \le \frac{\pi}{2}\, |\mathbf{b}_\perp|^2
\end{align}
with a Landau-Zener transition probability
$P_{\textrm{L-Z}}=\exp\left(-\frac{\pi}{2}
\frac{|\mathbf{b}_\perp|^2}{\hbar \dot{b_z}(t)}\right)$.  When
this condition is verified for all avoided level crossings on the
path $\Gamma_0$, the adiabatic theorem~\cite{Schiff68} may be
applied to the non-degenerate state $\alpha$ which time evolution
is approximatively
\begin{align}
\k{\psi_\alpha(t)}\approx
e^{-i\eta_\alpha(t)+i\gamma_\alpha(t)}\k{\alpha(\boldsymbol{\vartheta}(t))}\,,
\label{adiabatic}
\end{align}
where $\k{\alpha(\boldsymbol{\vartheta}(t))}$ is an instantaneous
eigenstate ($\hat{\mathcal{H}}\k{\alpha}=E_\alpha \k{\alpha}$),
the phase $\eta_\alpha(t)$ is the usual dynamical phase factor
\begin{align}
\eta_\alpha(t)=\frac{1}{\hbar}\int_0^tE_\alpha(t')\,dt'\,,
\end{align}
and $\gamma_\alpha(t)$, Berry's geometrical phase, was introduced in
last section.  In realistic systems, relaxation processes restrict the
use of TQCP to the ground state, and inelastic transitions to the first
excited state will be shown to dominate quantization errors.

\begin{figure}
\includegraphics[width=\columnwidth]{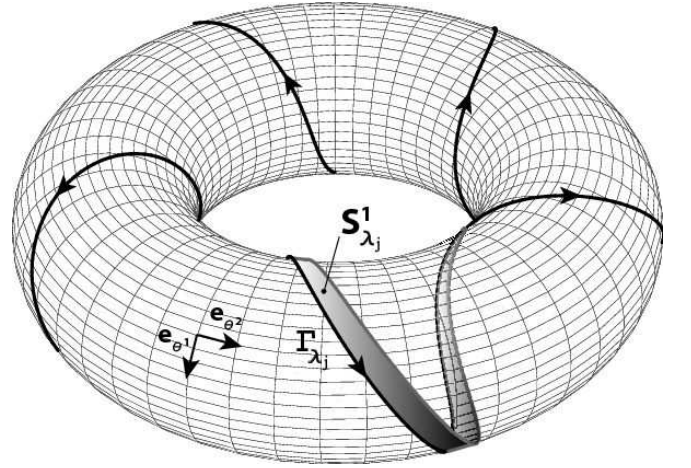}
\caption{Representation of a closed helical path $\Gamma_{\lambda_j}^c$ lying
on the torus $\mathbb{T}^2$ (here with a frequency ratio $\nu_{2}/\nu_{1}=5$).
The geometrical contribution $Q^{\textrm{geo}}$ of the charge $Q$ over a period
$T_2=\frac{2\pi}{\nu_2}$ is proportional to the flux of
$\mathbf{B}^{\band{\alpha}}$ through the infinitesimally thin strip
$\mathbf{S}^1_{\lambda_j}$.} \label{torus}
\end{figure}

Let us introduce the family of paths $\{\Gamma_\lambda\}$ on
$\mathbb{T}^2$ differing from $\Gamma_0$ by a shift of $\vartheta^2$ by
a constant angle $\lambda\in[0,2\pi)$.  In the next sections, the
helical family
\begin{align}
\Gamma_\lambda:t\in[0;T]\to\boldsymbol{\vartheta}_\lambda(t)=
\big(2\pi\,\nu_1\,t,2\pi\,\nu_2\,t+\lambda\big)
\label{chemins}
\end{align}
where the angles $\vartheta^1$ and $\vartheta^2$ rotate at frequencies $\nu_1$
and $\nu_2$, will be used in a practical implementation of TQCP.  When the
frequencies are commensurate, the paths $\{\Gamma_\lambda\}$ are closed. One of
them is represented pictorially in Fig.~\ref{torus}, having commensurate
frequencies $\nu_1=5\,\nu_2$. Each path $\Gamma_\lambda$ begins at the point
$\boldsymbol{\vartheta}_i=(0,\lambda)$ and the whole family
$\{\Gamma_\lambda\}$ covers entirely the torus as $\lambda$ is swept from 0 to
$2\pi$.

The quantity of interest is the value of ``the transferred charge''
$Q(\Gamma_\lambda)$ accumulated over the path $\{\Gamma_\lambda\}$ which is
covered in a period $T$. Integrating $\dot{Q}$ over time gives:
\begin{align}
Q(t)=\int_0^t\big\langle\partial_2\hat{\mathcal{H}}(\boldsymbol{\vartheta}(t'))\big\rangle\,dt'\,.
\label{integrand}
\end{align}
The integrand in Eq.~\ref{integrand} is split in two parts
\begin{align}
\partial_2\big\langle\hat{\mathcal{H}}\big\rangle
-2\Re e\,\b{\psi_\alpha}\hat{\mathcal{H}}\k{\partial_\varphi\psi_\alpha}.
\label{twocontrib}
\end{align}
Each term contributes to the transferred charge $Q(t)$: the first one leads to
a dynamical contribution $Q^{\rm dyn}$, while the second one specifies the
geometrical pumped charge $Q^{\rm geo}$. To identify the dynamical
contribution, we take the time derivative of the adiabatic evolution (Eq.
\ref{adiabatic}) and apply Schr\oe{}dinger equation
$i\hbar\,\k{\dot{\psi}_\alpha}=\hat{\mathcal{H}}\k{\psi_\alpha}$ to express
\begin{align}
\hat{\mathcal{H}}\k{\psi_\alpha}&=
e^{-i(\eta_\alpha-\gamma_\alpha)}\Big[i\hbar\,\k{\dot{\alpha}}+\big(E_\alpha-\hbar\,\dot{\gamma}_\alpha\big)\k{\alpha}\Big]\,.
\label{Hpsi}
\end{align}
The expectation value of $\mathcal H$ and its phase derivative follow
from Eq.~\ref{Hpsi} and Eq.~\ref{adiabatic},
\begin{align}
\label{dynamique}\partial_2\big\langle\hat{\mathcal{H}}\big\rangle&=\partial_2
E_\alpha+\hbar\,\partial_2\big(i\bk{\alpha}{\dot{\alpha}}-\dot{\gamma}_\alpha\big)\,.
\end{align}
Since $\dot{\gamma}_\alpha=i\bk{\alpha}{\dot{\alpha}}$, the last
two terms on the right hand side disappear. When integrated over
the period $T$ this first contribution $Q^{\rm
dyn}(\Gamma_\lambda)$ to the transferred charge is also the
derivative of the \textit{dynamical phase} with respect to the
initial angle $\lambda$
\begin{align}
\label{dynamic}
Q^{\rm dyn}(\Gamma_\lambda)=\int_0^{T}
\!\!\partial_2E_{\alpha}(\boldsymbol{\vartheta}_\lambda(t))\,dt=
\hbar\,\frac{d\eta_{\alpha}(\Gamma_\lambda)}{d\lambda}\,,
\end{align}
where the definition (Eq.~\ref{chemins}) of the helical paths has been used to
transform the partial derivative of the integrand into a total derivative of
the dynamical phase with respect to the initial angle $\lambda$. This quantity
is just the difference between the total accumulated dynamical phases on the
neighboring paths $\Gamma_{\lambda+d\lambda}$ and $\Gamma_{\lambda}$ normalized
to the angle increment $d\lambda$.

We now turn to the geometrical contribution $Q^{\rm geo}$, which comes
from the second term in Eq.~\ref{twocontrib}. Taking the $\vartheta^2$
derivative of  Eq.~\ref{adiabatic} yields
\begin{align}
\k{\partial_2\psi_\alpha}&=
e^{-i\big(\eta_\alpha-\gamma_\alpha\big)}\Big[\k{\partial_2\alpha}+
i\,\partial_2\big(\gamma_\alpha-\eta_\alpha\big)\k{\alpha}\Big]\,.
\label{phiderivative}
\end{align}
Using Eqs.~\ref{dynamique} and~\ref{phiderivative}, the scalar
product
$\b{\psi_\alpha}\hat{\mathcal{H}}\k{\partial_2\psi_\alpha}$ gives
several terms, but only one of them is not purely imaginary,
namely $-i\hbar\bk{\dot{\alpha}}{\partial_2\alpha}$. When
integrated over time, the result does not depend on the dynamics,
but only on the path geometry. Hence the \textit{geometric} pumped
charge $Q_\alpha^{\rm geo}(\Gamma_{\lambda})$ is
\begin{align}
Q^{\rm geo}(\Gamma_{\lambda})&=
-2\,\Re\int_0^{T}\b{\psi_\alpha(t)}\hat{\mathcal{H}}(\boldsymbol{\vartheta}_\lambda(t))\k{\partial_2\psi_\alpha(t)}\,dt\notag\\
&=\hbar\,\int_{\Gamma_{\lambda}}\!\!\!2\Im\,\bk{\partial_2\alpha(\boldsymbol{\vartheta}(t))}{d\alpha(\boldsymbol{\vartheta}(t))}\,.\label{geometricpart}
\end{align}
This charge can be expressed in term of a geometrical phase by
rewriting
\begin{align}
2\Im m\,\bk{\partial_2\alpha}{d\alpha}=
i\Big[d\bk{\alpha}{\partial_2\alpha}-\partial_2\bk{\alpha}{d\alpha}\Big].
\end{align}
The second term is recognized the $\vartheta^2$ derivative of the
connexion $\mathcal{A}_\alpha$ (defined in Eq.~\ref{vecpot}) which
integral over a closed path is Berry's phase. When integrating
over the path $\Gamma_\lambda$, the first term only contributes at
the endpoints
$\boldsymbol{\vartheta_i}=\boldsymbol{\vartheta}_\lambda(0)$ and
$\boldsymbol{\vartheta_f}=\boldsymbol{\vartheta}_\lambda(T)$,
giving
\begin{align}
Q^{\rm geo}(\Gamma_{\lambda})=
-\hbar\int_{\Gamma_\lambda}\!\!\!\!\!\!\partial_2\mathcal{A}^{\band{\alpha}}&
+i\hbar\Big[\bk{\alpha(\boldsymbol{\vartheta_f})}{\partial_2\alpha(\boldsymbol{\vartheta_f})}\notag\\
&\;\;\;\;\;-\bk{\alpha(\boldsymbol{\vartheta_i})}{\partial_2\alpha(\boldsymbol{\vartheta_i})}\Big]\,,
\label{qgeometric}
\end{align}
these last two contributions being essential to enforce the gauge
invariance of $Q^{\rm geo}(\Gamma_{\lambda})$, a measurable
quantity. When the path $\Gamma_{\lambda}^{\rm c}$ is closed, the
endpoints contributions cancel, and $Q^{\rm
geo}(\Gamma_{\lambda})$ is the integral of the $\vartheta^2$
derivative of the vector potential
\begin{align}
Q^{\rm geo}(\Gamma^{\rm c}_{\lambda})&=
-\hbar\oint_{\Gamma_\lambda}\!\!\!\!\partial_2\mathbf{\mathcal{A}^{\band{\alpha}}}\,.
\label{geo_closed}
\end{align}
For the helical path family $\{\Gamma_\lambda\}$, we showed in
Eq.~\ref{dynamic} how the dynamical transferred charge
$Q^{\textrm{dyn}}(\Gamma_\lambda)$ could be expressed as the total
derivative of the dynamical phase with respect to the initial
angle $\lambda$. The same argument can be used here
\textit{mutatis mutandis} to the geometrical transferred charge
\begin{align}
\label{Q_geo} Q^{\rm geo}(\Gamma^{\rm
c}_{\lambda})=-\hbar\,\frac{d}{d\lambda}\oint_{\Gamma^{\rm
c}_\lambda}\!\!\!\mathcal{A}^{\band{\alpha}}=-\hbar\,\frac{d\gamma_{\alpha}(\Gamma^{\rm
c}_{\lambda})}{d\lambda}\,.
\end{align}

\begin{figure}
\centering
\includegraphics[width=0.8\columnwidth]{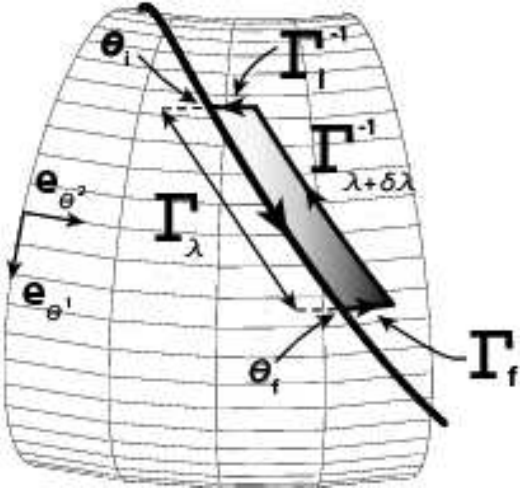}
\caption{The virtual path $\Sigma_\lambda$ is the sum of the four
path-segments $\Gamma_\lambda$, $\Gamma_f$,
$\Gamma_{\lambda+\delta\lambda}^{-1}$ and $\Gamma_i^{-1}$ lying on
the torus $\mathbb{T}^2$. The geometrical contribution
$Q^{\textrm{geo}}$ to the charge $Q$ over the segment
$\Gamma_\lambda$ is proportional to the circulation of
$\mathbf{A}^{\band{\alpha}}$ along the ``virtual'' closed path
$\Sigma_\lambda$.} \label{virtual}
\end{figure}

This formula presents the advantage to be easily {\em generalized to
open paths}  thanks to the endpoint contributions in
Eq.~\ref{qgeometric}.  In a first step, the integral of the angle
derivative
\begin{align}
- \hbar\int_{\Gamma_{\lambda}}\!\!\!\!\partial_2
\mathcal{A}^{\band{\alpha}}= \frac{\hbar}{\delta\lambda}\bigg[
\int_{\Gamma_{\lambda}}+\int_{\Gamma^{-1}_{\lambda+\delta\lambda}}\bigg]\mathcal{A}^{\band{\alpha}}\,,
\label{remark}
\end{align}
is rewritten as a difference between two paths shifted by the
infinitesimal $\delta \lambda$, which becomes a sum when one of
the segment is integrated in the opposite direction
($\Gamma^{-1}_{\lambda+\delta\lambda}$). These two paths can be
connected by infinitesimal vertical segments $\Gamma^{-1}_i$ and
$\Gamma_f$ at their endpoints $\boldsymbol{\vartheta}_i$ and
$\boldsymbol{\vartheta}_f$ as shown in Fig.~\ref{virtual}.  The
endpoints contributions in Eq.~\ref{qgeometric} can be rewritten
as line integral of the vector potential over these end-segments
as
\begin{align}
\label{findebut}
\left\{\begin{array}{rl}i\hbar\,\bk{\alpha(\boldsymbol{\vartheta}_f)}{\partial_2\alpha(\boldsymbol{\vartheta}_f)}&=
\displaystyle\frac{\hbar}{\delta\lambda}\int_{\Gamma_f}\!\!\mathcal{A}^{\band{\alpha}}\\
 & \\
-i\hbar\,\bk{\alpha(\boldsymbol{\vartheta}_i)}{\partial_2\alpha(\boldsymbol{\vartheta}_i)}&=
\displaystyle\frac{\hbar}{\delta\lambda}\int_{\Gamma_i^{-1}}\!\!\mathcal{A}^{\band{\alpha}}\end{array}\right..
\end{align}
When combining the four path-segments together, a closed path
$\Sigma_\lambda$ is constructed from the path $\Gamma_\lambda$ which is
one of its line-segments as drawn in Fig.~\ref{virtual}. On this closed
path $\Sigma_\lambda$, the integral of the vector potential becomes
precisely Berry's phase. By constructing the four segments {\em virtual
path} $\Sigma_\lambda$, one of which is the physical path
$\Gamma_\lambda$ of interest, the geometrical transferred charge on
$\Gamma_\lambda$ can be written as
\begin{align}
Q^{\rm geo}(\Gamma_\lambda)=\hbar\,
\frac{\gamma_\alpha(\Sigma_\lambda)}{\delta\lambda}\,. \label{Q_open}
\end{align}

In contrast with Eq.~\ref{Q_geo} which gives only a global description of the
geometric charge on a closed path, this expression for the pumped charge can be
used on any arbitrary paths.  They are relevant if noise or error in the
control of parameters exist.

What is the benefit of this formulation in term of Berry's phase ?
One is practical: Berry's phase can be computed efficiently.   The
gauge dependence of the vector potential
$\mathbf{A}^{\band{\alpha}}$ introduces a difficulty which can be
circumvented in two ways.  Berry's phase can be computed as the
flux of the magnetic induction $\mathbf{B}^{\band{\alpha}}$ using
Eq.~\ref{flux-formula}.  The two dimensional integration can
however be tedious to compute, particularly when the surface is
warped. Alternatively, King-Smith, Vanderbilt and
Resta~\cite{King-Smith93,Resta94} formulated Berry's phase in term
of a gauge invariant expression by discretizing the
one-dimensional path $\Gamma_\lambda$.  Let
$\boldsymbol{\vartheta}_j$ be N points on $\Sigma_\lambda$
splitting it in $N$ small segments. The line integral of the
vector potential can then be expressed as the
invariant\cite{Bargmann64}
\begin{align}
\gamma_{\alpha}(\Sigma_\lambda)\approx
-\arg\prod_{j=0}^{N-1}\bk{\alpha(\boldsymbol{\vartheta}_j)}{\alpha(\boldsymbol{\vartheta}_{j+1})}\,.
\label{Resta}
\end{align}
In this way, any local gauge change cancels out between bras and kets which
come each in pairs. It is easy to implement over complex paths and very
accurate. When shrinking the path $\Gamma_\lambda$ to an infinitesimal segment
between $\boldsymbol{\vartheta}_\lambda(t)$ and
$\boldsymbol{\vartheta}_\lambda(t+dt)$, one also gets the instantaneous
geometrical pumped charge:
\begin{align}
\delta Q^{\rm geo}(t)\approx
-\frac{\hbar}{\delta\lambda}\,\arg\prod_{j=0}^{3}\bk{\alpha(\boldsymbol{\vartheta}_j)}
{\alpha(\boldsymbol{\vartheta}_{j+1})}\,,
\label{Q_instantaneous}
\end{align}
where the four points are the extremities of the infinitesimal
paths (see Fig.~\ref{virtual}).  The geometrical pumped charge
$Q$, which is a physical quantity, can be tracked and measured
anywhere along any real path. Equations~\ref{Q_open}
and~\ref{Q_instantaneous} are thus of great practical value since
the local physical processes and experimental sources of errors in
the path can be analyzed on the quantity of interest $Q^{\rm
geo}$.

For particular sets of paths (the ``controlled paths''), the ``charge'' ${\dot
Q}=\big\langle\partial_2\hat{\mathcal{H}}\big\rangle$ can be quantized through
its relation to the Chern index of a closed surface for the ground state
eigenvector bundle.  Any closed surface containing one or more singularities
can in principle be used. For simplicity, the entire torus $\mathbb{T}^2$ will
be used here. There are several ways one can generate this two-dimensional
surface using a one dimensional path. The helical family $\{\Gamma_\lambda\}$,
is one of the possible families of controlled paths which generates the surface
$\mathbb{T}^2$. When the angular frequencies of $\vartheta^1$ and $\vartheta^2$
are commensurate $\nu_2=p\,\nu_1$, the angle $\vartheta^1$ winds $p$ times
around in a $\vartheta^2$ period. When $p$ is large, the helix covers densely
the torus.  Alternatively, the initial angle $\lambda$ can be swept from $0$ to
$2\pi$ to sweep the helix on the torus surface $p$-times.  Using this averaging
procedure, the dynamical contribution to the pumped charge averages out to
zero,
\begin{align}
\langle Q^{\rm dyn}\rangle&=
\frac{\hbar}{2\pi}\int_0^{2\pi}\frac{d\eta_\alpha(\Gamma^c_\lambda)}{d\lambda}\,d\lambda\\\notag
&=\hbar\Big(\eta_\alpha(\Gamma_{2\pi}^c)-\eta_\alpha(\Gamma_0^c)\Big)=0\,,
\end{align}
since $\Gamma_{2\pi}^c\equiv\Gamma_0^c$. When discussing the
geometrical contribution $\langle Q^{\rm dyn}\rangle$, it is
simpler, to split the helix into $p$ one-turn segments
$\Gamma_\lambda^{\rm 1~turn}$ (which are open paths). As the
initial angle $\lambda$ of the helix is swept from $0$ to $2\pi$,
each one-turn segment $\Gamma_\lambda^{\rm 1~turn}$ sweeps the
torus surface just once ($p$ times for the whole helix). For this
reason, it is simplest to compute the average of the geometric
charge over a $2\pi$ $\lambda$-period for this one turn segment
$\Gamma_\lambda^{\rm 1~turn}$ and multiply the result by $p$ for
the whole helix.  The integral over $\lambda$ can be made by
dividing the $2\pi$ period in $N$ small slices indexed by $j$ of
width $\Delta\lambda=\frac{2\pi}{N}$.

The contribution to the pumped charge over the one turn segment
$\Gamma_{\lambda_j}^{\rm 1~turn}$  for the $j^{\rm th}$ slice of
width $\Delta\lambda$ defines the helix strip of surface
$S_{\lambda_j}^1$ represented in Fig.~\ref{torus}.  Its boundary
is nothing but the virtual path $\Sigma_j^{\rm 1~turn}$ associated
to the one-turn segment $\Gamma_{\lambda_j}^{\rm 1~turn}$ (see
Fig.~\ref{virtual}).  Using Eq.~\ref{Q_open}, the pumped charge
averaged over this interval is
\begin{align}
\Delta Q^{\rm geo}_\alpha(\Gamma^{\rm
1~turn}_{\lambda_j})&=\frac{\hbar}{\Delta\lambda}\,
\gamma_\alpha\big(\Sigma_j^{\rm 1~turn} \big)= \\
\frac{\hbar}{\Delta\lambda}\oint_{\Sigma_j^{\rm 1~turn}}\!\!\!\!\!
\mathbf{A}^{\band{\alpha}}\cdot
d\mathbf{R}&=\frac{\hbar}{\Delta\lambda}
\iint_{S_{\lambda_j}^1}\mathbf{B}^{\band{\alpha}} \cdot {\hat n}\,
d\mathcal{S}\,, \nonumber
\end{align}
where Stokes theorem was used to express Berry's phase along the
virtual path $\Sigma_{\lambda}^{\rm 1~turn}$ as the flux of
$\mathbf{B}^{\band{\alpha}}$ through $S_{\lambda_j}^1$. When
summing over all the $j$ slices of height $\Delta\lambda$, these
elementary surfaces add up to the entire surface of the torus.
Hence, when averaged over $\lambda$, the geometrical charge
transferred becomes
\begin{align}
\langle Q^{\rm
geo}_\alpha\rangle&=\lim_{N\to\infty}\frac{1}{N}\sum_{j=1}^N
Q^{\rm geo}_{\alpha}(\Gamma^{\rm
1~turn}_{\lambda_j}) \nonumber \\
&=\frac{\hbar}{2\pi} \sum_{j=1}^N
\gamma_\alpha(\Sigma_{\lambda_j})
=\frac{\hbar}{2\pi}\oiint_{\mathbb{T}^2}
\mathbf{B}^{\band{\alpha}}\cdot{\hat n}\, d\mathcal{S}\,.
\end{align}
This last term is precisely the Chern index $c_1^{\band{\alpha}}$
of the surface $\mathbb{T}^2$ for the $\alpha$-eigenvector bundle.
For the whole helix, each one-turn segment contributes equally and
\begin{align}
\langle Q^{\rm
geo}\rangle=p\,\hbar\,c_1^{\band{\alpha}}(\mathbb{T}^2)\,.
\end{align}
The average of $Q^{\rm geo}$ over the family $\{\Gamma_\lambda^c\}$ is
quantized by the winding number $p$, and the Chern index of the torus
$\mathbb{T}^2$ with respect to the band $\alpha$. It is nonzero only if
degeneracies involving the band $\alpha$ are present inside the torus. As was
pointed out by Goryo and Kohmoto\cite{Goryo07}, invariances of the Hamiltonian
under mirror symmetries
$(\vartheta_1,\vartheta_2)\to\{(-\vartheta_1,\vartheta_2)$ or
$(\vartheta_1,-\vartheta_2)\}$ (i.e.
$\hat{\mathcal{H}}(\vartheta_1,\vartheta_2)=\hat{\mathcal{H}}(-\vartheta_1,\vartheta_2)$
or $\hat{\mathcal{H}}(\vartheta_1,-\vartheta_2))$ and ``time-reversal''
symmetry $\boldsymbol{\vartheta}\to-\boldsymbol{\vartheta}$ (i.e.
$\hat{\mathcal{H}}(\boldsymbol{\vartheta})=\hat{\mathcal{H}}^*(-\boldsymbol{\vartheta}))$
are incompatible with a nonzero Chern index. This is because
$\mathbf{B}^{\band{\alpha}}$ is an axial vector: the mirror symmetry
$(\vartheta_1,\vartheta_2)\to(\vartheta_1,-\vartheta_2)$ leaves the torus
invariant, but $\mathbf{B}^{\band{\alpha}}$ changes sign with respect to the
local natural basis
$(\mathbf{e}_{\vartheta^1},\mathbf{e}_{\vartheta^2},\hat{n})$ ($\hat{n}$ is the
vector normal to the surface). Hence the mirror symmetry switches the sign of
$\mathbf{B}^{\band{\alpha}}(\vartheta^1,\vartheta^2)\cdot \hat{n}$ and the
integral of $\mathbf{B}^{\band{\alpha}}\cdot \hat{n}$ over the torus vanishes.
This property can be used locally to detect the presence of singularities in
the eigenvector bundle.  For example, in the spin representation
(Eq.~\ref{Zeeman}) close to a singularity, under the mirror symmetry
($b_z\rightarrow -b_z$) the $\k{+}$ and $\k{-}$ are mapped into each other and
each eigenvector bundle is not preserved separately.  The same behavior occurs
under time-reversal.

Goryo and Kohmoto\cite{Goryo07} generalized the relation between the
expectation value of a derivative of the Hamiltonian and the Chern indices on
D-dimensional torii, with application to a number of problems (IQHE in 2D and
3D dimensions, ACJE, etc.). In these problems, the averaging over the whole
torus can be made directly, but in our case the physical quantity $Q$ is
generated by paths: a path description cannot be avoided. Since the average
$\langle Q\rangle$ over a family of commensurate paths $\{\Gamma^c_\lambda\}$
is quantized, the value of $Q(\Gamma^c_\lambda)$  for a given $\lambda$
fluctuates around the integer mean value.   It is interesting to know how these
fluctuations decrease with winding number. Since the torus is covered densely
at large $p$, we expect a more accurate quantization as the winding number $p$
get larger, irrespective of the value of $\lambda$.   A more accurate averaging
of the dynamical charge improves the quantization.   This will be easiest if
$\vartheta^2$-dependence of the energy $E_\alpha(\vartheta^2)$ is weak since
$Q^{\rm dyn}\sim\partial_2E_\alpha$ . TQCP is an asymptotic quantization, which
works best for the ground state which is  most robust against incoherent
processes.  For a two-dimensional torus, a number of paths can be chosen, the
only requirement for TQCP being the $\vartheta^2$ periodicity.

Next section, devoted to the Cooper Pair Pump is a physical example where TQCP
can be implemented concretely.

\section{\label{sec:CPP}Topological properties of the Cooper Pair Pump}
\begin{figure}[t]
\centering
\includegraphics[width=0.9\columnwidth]{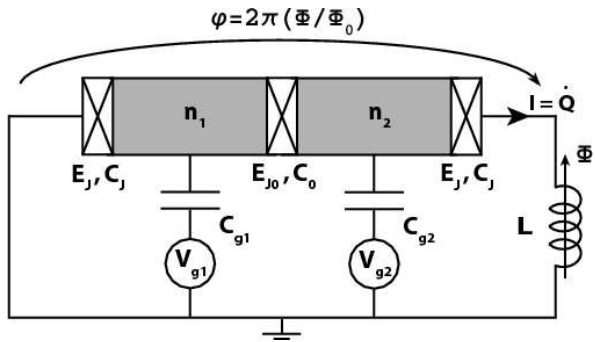}
\caption{The circuit is made of a Cooper Pairs Pump closed on a small
inductance $L$, threaded by a magnetic flux $\Phi$. The charge on the
islands 1 and 2 can be tuned by two gate voltages $V_{g1}$ and
$V_{g2}$.  The inductance is used to bias the phase across the CPP,
$\varphi=2\pi\Phi/\Phi_0$.}\label{CPP}
\end{figure}

One of the simplest implementation for a Cooper pair pump (CPP) using a
superconducting circuit is represented on Fig.~\ref{CPP}. Phase biasing
is achieved by closing the CPP on a small inductance $L$, threaded by a
magnetic flux $\Phi$. Its magnetic contribution to the energy is
$\frac{1}{2L}({\hat \phi} - \varphi)^2$ where $\hat \phi$ is the phase
difference across $L$. For small $L$, it has a deep minimum at
$\varphi=2\pi\frac{\Phi}{\Phi_0}$: this inductance and the CPP series
capacitance $C_s$ form an harmonic oscillator which frequency
$(2\pi\sqrt{LC_s})^{-1}$ exceeds all other energies, effectively
blocking the quantum variable $\hat \phi$ at the value $\varphi$ (the
center of the ground state wavefunction). $\varphi$ is then a parameter
tunable by the magnetic flux $\Phi$.

\begin{figure}[b]
\centering
\includegraphics[width=0.9\columnwidth]{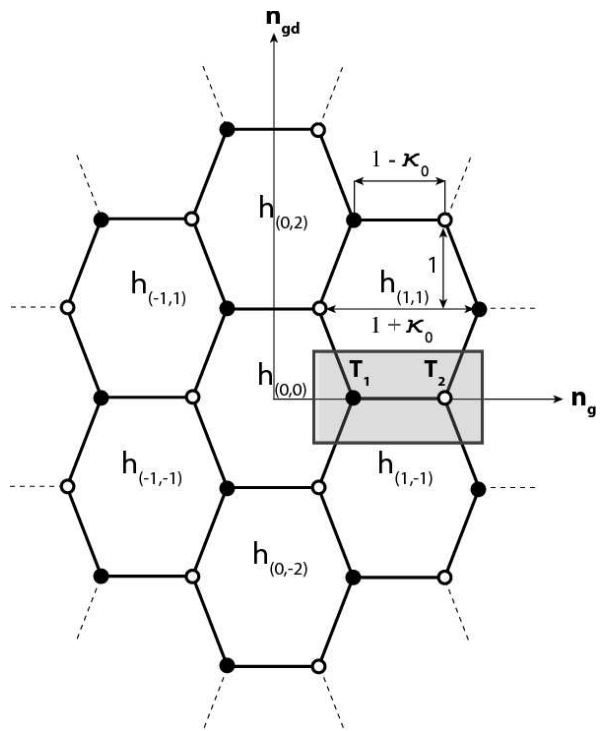}
\caption{Stability diagram for the charging Hamiltonian (Eq.~\ref{HC})
in the $n_{gs}\!-n_{gd}$ plane.  The charging energy is minimized
inside the hexagonal areas $h_{(n_s,n_d)}$ shown.  The boundaries
between hexagons are line of degeneracies between charge states, while
the vertices are points of triple degeneracies.  The two vertices $T_1$
and $T_2$ form the unit cell for this hexagonal lattice. The
coordinates of $T_1$ are $(\frac{1+\kappa_0}{2}, 0)$. Also shown are
the topological charges of the lattice of degeneracies in the plane
$\varphi=\pi$ (see text).} \label{honeycomb}
\end{figure}

\begin{figure}
\begin{center}
\includegraphics[width=\columnwidth]{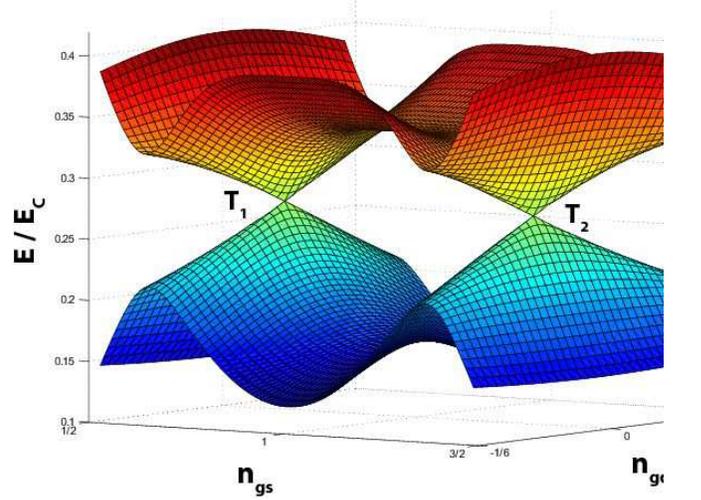}
\end{center}
\caption{(a): The two lowest energy manifolds computed in the
rectangular area of the $n_{gs}\!-n_{gd}$ plane shown in
Fig.~\ref{honeycomb} for the constant phase $\varphi=\pi$.  In the
vicinity of the isolated degeneracies $\mathbf{T_1}$ and
$\mathbf{T_2}$, the energy sheets form a conical intersection,
referred to as ``diabolical points'' (same shape as a diabolo). At
these points there are topological singularities in the bundle
$\mathcal{F}^{\band{-}}$ (Eq.~\ref{bundle}).}\label{diabolical}
\end{figure}

The three Josephson junctions, with small capacitances define two
superconducting islands with sufficient large electrostatic
energies to limit charge fluctuations through the junctions.  Let
$n_{i}$ be the excess number of Cooper pairs (with respect to
charge neutrality) on island $i$. The electrostatic energies of
each island can be tuned independently using a gate voltage
$V_{g_i}$ through the gate capacitances $C_{g_i}$.  The induced
charge polarization on the island $i$ is
$n_{g_i}=C_{g_i}V_{g_i}/(2e)$ in units of $2e$.  For convenience,
we use the total charge $n_s=n_1+n_2$ on the double-island and the
charge asymmetry $n_d=n_1-n_2$ between them as the natural basis
of charge states $\{\k{n_s,n_d}\}$. Taking the two external
junctions with the same Josephson energy $E_{J}$ and capacitance
$C_J$, and ($E_{J0}$, $C_0$) for the central junction Josephson
energy and capacitance, the charging energy of the CPP reads
\begin{equation}
\hat{\mathcal{H}}_C=E_C\Big[\big(\hat{n}_{s}-n_{gs}\big)^2+\kappa_0\big(\hat{n}_{d}-n_{gd}\big)^2\Big]\,,
\label{HC}
\end{equation}
where $E_C=\frac{(2e)^2}{4C_J}$ is the Coulomb energy,
$n_{gs}\!\!=n_{g_1}\!\!+n_{g_2}$, $n_{gd}\!\!=n_{g_1}\!\!-n_{g_2}$ and
$\kappa_0=\frac{C_J}{2C_0+C_J}$ is a capacitance ratio (of order
$\frac{1}{3}$). In addition to the phase bias $\varphi$, the induced charge $n_
{gs}, n_{gd}$ are tunable parameters of the Hamiltonian: the parameter space
$\mathbb{P}$ is here three dimensional, and a point $\mathbf{R}$ in
$\mathbb{P}$ is specified by its coordinates $(n_{gs}, n_{gd}, \varphi)$. One
easily checks that the charge state $\k{n_s,n_d}$ is the ground state which
minimizes the parabolas in $\hat{\mathcal{H}}_C$ (Eq.~\ref{HC}) inside the
hexagonal area $h_{(n_s,n_d)}$ centered at the point
$(n_{gs}\!\!=n_s,n_{gd}\!\!=n_d)$, in the $n_{gs}\!-n_{gd}$ plane
(Fig.~\ref{honeycomb}).  On the line boundaries between hexagons, two
electrostatic states have the same energies while the vertices are points of
triple degeneracies.  This hexagonal lattice of triple degeneracies has two
points in its unit cell, chosen here as
$\{T_1\}=\left(n_{gs}=\frac{1+\kappa_0}{2},n_{gd}=0\right)$ and
$\{T_2\}=\left(n_{gs}=\frac{1-\kappa_0}{2}, n_{gd}=1\right)$. The Josephson
tunnelling, which ``translates'' Cooper pairs across the junctions, can be
expressed in term of the variables conjugate to the total charge ${\hat n}_s$
and charge asymmetry ${\hat n}_d$, ${\hat \Theta}_s$ and ${\hat \Theta}_d$
which are the generators of charge translations,
\begin{equation}
\hat{\mathcal{H}}_J=-2E_J\cos\hat{\Theta}_s\cos(\hat{\Theta}_d-
\kappa_J\,\varphi)-E_{J0}\cos(2\hat{\Theta}_d+\kappa_0\,\varphi)\,,\label{HJ}
\end{equation}
where $\kappa_J=\frac{C_0}{2C_0+C_J}=\frac{1-\kappa_0}{2}$ is the
other capacitance ratio (also of order $\frac{1}{3}$). Since
$\hat{\mathcal{H}}_J$ delocalizes Cooper pairs, the charge states
are no longer eigenstates of the full Hamiltonian
$\hat{\mathcal{H}}=\hat{\mathcal{H}}_C+{\mathcal H}_J$ and the
degeneracies along the boundaries of the honeycomb lattice are
lifted. Nevertheless, if $E_J\approx E_{J0}\le E_c$, accidental
isolated degeneracies persist in $\mathbb{P}$ in the vicinity of
the points $\mathbf{T_1}=\{T_1, \varphi=\pi\}$ and
$\mathbf{T_2}=\{T_2,\pi\}$ and all their equivalents under lattice
translations in $\mathbb{P}$. In the special case where
$E_J=E_{J0}$ (homogenous array), the degeneracies are placed at
$\mathbf{T_1}$ and $\mathbf{T_2}$, and shift along the
$n_{gs}$-axis for asymmetric arrays. As an illustration, the
energy manifolds for the two lowest levels $\k{\pm}$, represented
in Fig~\ref{diabolical}, show the two conical intersections in the
points $\mathbf{T_1}$ and $\mathbf{T_2}$ in $\mathbb{P}$ for
$E_J=E_{J0}$. Symmetries of the total Hamiltonian are most
explicit after the unitary transformation generated by
$U(\varphi)=e^{-i\kappa_J\varphi\,\hat{n}_d}$, which leaves the
charging Hamiltonian and shifts $\hat{\mathcal{H}}_J$ into
\begin{align}
\widetilde{\mathcal{H}}_J&=U(\varphi)\,\hat{\mathcal{H}}_J\,U^{\dag}(\varphi)\notag\\
&=-2E_J\cos\hat{\Theta}_s\cos\hat{\Theta}_d-E_{J0}\cos(2\hat{\Theta}_d+\varphi).
\end{align}
In this representation, the phase bias appears across the central junction
instead of being distributed across the three junctions according to the
electrostatic voltage drop.  The mirror symmetry $n_{gs}\rightarrow -n_{gs}$
keeps the physics unchanged while $(\hat{n}_s, \hat{\theta}_s)\to (-\hat{n}_s,
-\hat{\theta}_s)$. When the phase $\varphi$ is equal to $0$ or $\pi$, the
Hamiltonian is real and also invariant under a second mirror symmetry
$n_{gd}\rightarrow -n_{gd}$, while $(\hat{n}_d,\hat{\Theta}_d)\to
(-\hat{n}_d,-\hat{\Theta}_d)$. The integer translations on the honeycomb
lattice $(n_{gs},n_{gd})\to(n_{gs}+n_{s0},n_{gd}+n_{d0})$ induced by the
translation operator $e^{-i(n_{s0}\hat{\Theta}_s+n_{d0}\hat{\Theta}_d)}$ leads
to different numbers of Cooper pairs on the island $(n_s-n_{s0},n_d-n_{d0})$.
These states are equivalent but physically distinguishable. Finally, the
Hamiltonian is $2\pi$-periodic in $\varphi$ and phases differing by multiple of
$2\pi$ leads to identical physical states: $\varphi$ plays the same role here
as the $\vartheta^2$ variable in the preceding section.

\begin{figure}
\centering
\includegraphics[width=\columnwidth]{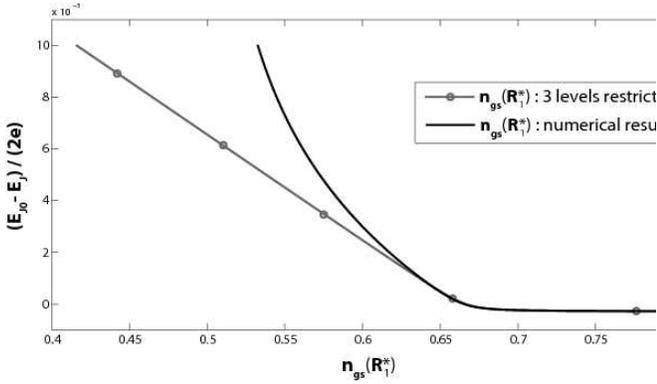}
\caption{Comparison of position of the degeneracy point $\mathbf{R_1}^*$
between the analytic model described in the text and the exact numerical
result.} \label{comparaison}
\end{figure}

In the same fashion as in section \ref{sec:fiber-bundle}, let us
construct explicitly the two-levels approximation of
$\hat{\mathcal{H}}$ in the vicinity of the degeneracy point
$\mathbf{T_1}$. Since circuit asymmetries do not affect the
topology of the eigenvector bundle, it is simpler to take
symmetric junctions $E_J=E_{J0}$ and assume that the ratio
$\beta=\frac{E_J}{2E_C}$ between Josephson and charging energies
remains small.  Writing the small deviations from the triple point
$\mathbf{T_1}$ as $\sigma=n_{gs}-\frac{2}{3}$, $\delta=n_{gd}$,
$\psi=\varphi-\pi$, the projection of the Hamiltonian
$\hat{\mathcal{H}}$ on the basis of charge states
$\{\k{0,0}',\k{1,1}',\k{1,-1}'\}$,
($\k{n_s,n_d}'=U(\varphi)\k{n_s,n_d}$), is represented by the
matrix
\begin{align}
\hat{\mathcal{H}}=E_C\left(\begin{array}{ccc} \frac{4\sigma}{3} & -\beta & -\beta  \\
-\beta & -2\frac{(\sigma+\delta)}{3} & \beta(1-i\psi) \\
-\beta  & \beta(1+i\psi) & -2\frac{(\sigma-\delta)}{3}
\end{array}\right),
\end{align}
to first order in the deviation
$\delta\mathbf{R}=(\sigma,\delta,\psi)$. At $\mathbf{T_1}$
($\sigma=\delta=\psi=0$), the two lowest eigenstates
\begin{align}
\k{+}&=\frac{1}{\sqrt{3}}\bigg(\k{0,0}'-\frac{\sqrt{3}-1}{2}\k{1,1}'+\frac{\sqrt{3}+1}{2}\k{1,-1}'
\bigg), \nonumber\\
\k{-}&=\frac{1}{\sqrt{3}}\bigg(\k{0,0}'+\frac{\sqrt{3}+1}{2}\k{1,1}'-\frac{\sqrt{3}-1}{2}\k{1,-1}'
\bigg) \nonumber
\end{align}
are degenerate with energy $-\frac{E_J}{2}$ (ground states), and the first
excited state
\begin{equation}
\k{e(\mathbf{T_1})}=\frac{1}{\sqrt{3}}\Big(\k{0,0}'-\k{1,1}'-\k{1,-1}'
\Big)
\end{equation}
has $E_J$ for eigenvalue.  As discussed in Sec.~\ref{sec:fiber-bundle}, an
isotropic spin representation of the Hamiltonian in the $\k{\pm}$ subspace
require a deformation of the parameter space $\mathbb{P}$ represented by the
matrix,
\begin{align}
\hat{C}=\left(\begin{array}{ccc}\frac{4}{3}\,E_C&0&0\\
0&0&-\frac{1}{\sqrt{3}}\,E_J\\
0&\frac{4}{3\sqrt{3}}\,E_C&0\end{array}\right)
\end{align}
which amounts here to a symmetry (the flip of the $\delta$ and $\psi$
axes changes the space orientation) and a linear deformation.  This
transformation specifies the effective magnetic field
($\mathbf{b}={\hat C}\delta\mathbf{R}$),
$b^x=\frac{4}{3}\,E_C\,\sigma$, $b^y=-\frac{1}{\sqrt{3}}\,E_J\psi$ and
$b^z=\frac{4E_C}{3\sqrt{3}}\,\delta$, such that the projection of the
Hamiltonian on the $\k{\pm}$ degenerate subspace reduces to a
spin-$\frac{1}{2}$ Hamiltonian
\begin{align}
\hat{\mathcal{H}}_{\pm (\mathbf{T_1})}&=
\frac{1}{2}\left(\begin{array}{cc} b^z & b^x-i\,b^y\\
b^x+i\,b^y & -b^z
\end{array}\right)=\frac{1}{2}\,\boldsymbol{\sigma}\cdot\mathbf{b}(\mathbf{R}).
\end{align}
The two lowest levels have a conical intersection at the
degeneracy point $\mathbf{T_1}$
\begin{align}
E_\pm(\mathbf{b})=\pm\, \frac{|\mathbf{b}|}{2}\,.
\end{align}
Following the discussion in the Sec.~\ref{sec:fiber-bundle}, the topological
charge in the spin representation and in the original parameter space are
identical up to the sign of the determinant of $\hat C$ which is positive.
Hence the topological charge of the ground state is
$q^{\band{0}}(\mathbf{T_1})=+1$ and $-1$ for the first excited band. Using the
same arguments, the topological charge at the degeneracy $\mathbf{T_2}$ is
$q^{\band{0}}(\mathbf{T_2})=-1$. Similarly all degeneracies obtained by lattice
translation from $\mathbf{T_1}$ (resp. $\mathbf{T_2}$) have a topological
charge of $+1$ (resp. $-1$) for the ground state.  As mentioned in
Sec.~\ref{sec:fiber-bundle}, the transformation properties of the eigenstates
bundle under mirror and time-reversal symmetries allow to detect the presence
of a degeneracy locally (in the spin-$\frac{1}{2}$ representation). Here, the
$\k{\pm}$ states map into each other, and each eigenvector bundle is not
preserved separately by these transformations.  When $E_{J0}$ deviates from
$E_J$, the degeneracies move continuously away from $\mathbf{T_1}$ and
$\mathbf{T_2}$. Using the same $3\times 3$ matrix representation for
$\hat{\mathcal{H}}$, the degeneracies slide along the $n_{gs}$ axis in the
vicinity of $\mathbf{T_1}$ and $\mathbf{T_2}$ as
\begin{align}
\mathbf{R}_{1,2}^*=\Bigg(n_{gs}(\mathbf{T_1})\mp\displaystyle\frac{1}{4E_C}\bigg[E_{J0}-\frac{E_J^2}{E_{J0}}\bigg],0,\pi\Bigg)\,,
\end{align}
for small deviations $|E_{J0}-E_{J}|$. In this shift, the degeneracies keep
their topological charge,
$q^{\band{0}}(\mathbf{R}_{i}^*)=q^{\band{0}}(\mathbf{T}_i)$. In
Fig.~\ref{comparaison}, the analytic and exact positions of the degeneracy
points are compared: the agreement deviates rapidly as one moves away from
$\mathbf{T_1}$.

The presence of degeneracies in the lowest band of the CPP allows to quantize
the pumped current opening accurate application for metrology, the topic of
next section.

\section{\label{sec:charge-quantization}Quantization in the Cooper Pair Pump}
\label{CQ}

The idea of using single electron pumps\cite{Lafarge93} as current
standard dates from the early 90's.  The original circuits uses normal
island separated by tunnel junction were biased by a small dc voltage
$V_B$. If the two gates voltages are driven in quadrature, the systems
undergoes a circular cycle centered around point $\mathbf{T_1}$. The
electrostatic ground state changes cyclically
($\k{0,0}\rightarrow\k{1,-1}\rightarrow\k{1,1}\rightarrow\k{0,0}$) as
one crosses one of the three degeneracy lines intersecting at $T_1$.
After one cycle, a single charge is transferred through the electron
pump. If the cycles are sufficiently slow, the charge relaxation (e.g.
$\k{0,0}\rightarrow\k{1,-1}$) is inelastic but has sufficient time to
complete. Since the process is stochastic, errors occur and limit the
accuracies of normal electron pumps. Also, the timescale for charge
relaxation are typically of order $\tau=(R_T C)^{-1}$, where $R_T$ is
the tunnelling resistance.  For realistic circuits, $\tau$ rarely
exceeds $10^{-6}$ sec., and pumped currents do not exceed a few
pico-Amp\`ere.

\begin{figure}[h]
\centering
\includegraphics[width=\columnwidth]{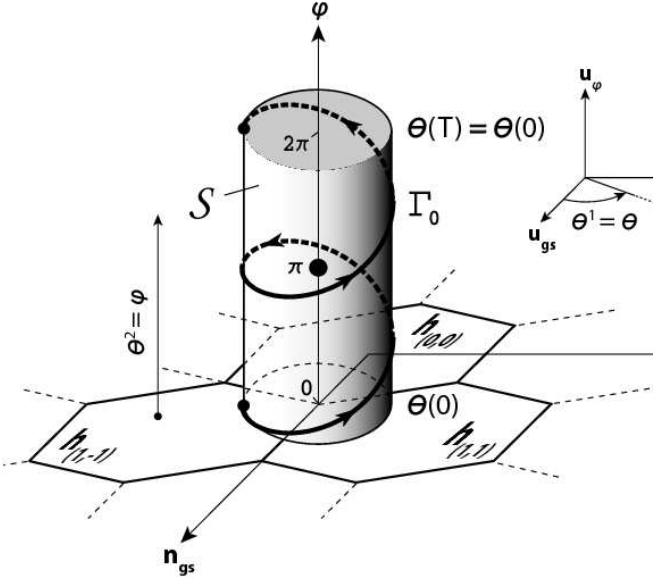}
\caption{ A helical closed path $\Gamma_0$ in $\mathbb{P}$ on the surface of a
cylinder enclosing the degeneracy $\mathbf{T_1}$. Since $\varphi$ is cyclic,
the surface $\mathcal{S}$ is a two-dimensional torus and $\Gamma_0$ is closed
for integer frequency ratio $p=\nu_{\theta}/\nu_{\varphi}$ (here
$2$).}\label{Fig-path}
\end{figure}

This is on of the motivation for studying Cooper-pair
pumps\cite{Zorin96,Bibow02,Niskanen03} (CPP) to circumvent the
stochasticity of normal electron devices.  Here, we show that the
charge transferred can be quantized topologically by using {\em
controlled paths} in parameter space (TQCP). The CPP's circuit
delivers a current $I$ which is equal to the charge transferred
$Q$ per unit of time : $I=\dot{Q}$ (see Fig.~\ref{CPP}). Let's
return to Eqs.~\ref{HC} and~\ref{HJ} and consider that
$\hat{\varphi}$ is still a quantum degree of freedom conjugated to
a charge operator $\hat{q}$, i.e. $[\hat{q},\hat{\varphi}]=i$. The
time evolution of the mean value of $\hat{q}$ is
$\frac{d\langle\hat{q}\rangle}{dt}=-\frac{i}{\hbar}\langle[\hat{q},\hat{\mathcal{H}}]\rangle$,
and is equal to
$\frac{1}{\hbar}\,\langle\partial_{\hat{\varphi}}\hat{\mathcal{H}}\rangle$.
Since the small inductance blocks the quantum fluctuation in
$\hat{\varphi}$, it can be taken as classic and the pumped current
is
\begin{align}
I=\dot{Q}=\frac{2e}{\hbar}\,\langle\partial_{\varphi}\hat{\mathcal{H}}\rangle\,.
\end{align}
Since $\varphi$ and $\vartheta^2$ have the same $2\pi$ periodicity
$\frac{\hbar}{2e} I$ has the exact expression (Eq.~\ref{formedebase}) as
required for the TQCP procedure discussed in Sec.~\ref{sec:TQCP}.  Consider now
the cylinder $\mathcal{S}$ in parameter space represented in
Fig.~\ref{Fig-path}, which axis lies in the $\varphi$ direction. Its section in
the $n_{gs}$-$n_{gd}$ plane has a radius $\rho$ of order
$\frac{1-\kappa_0}{2}\simeq\frac{1}{3}$ and its height on the $\varphi$-axis is
$2\pi$.  Since the end-faces $\varphi=0$ and $\varphi=2\pi$ are physically
equivalent, this cylinder $\mathcal{S}$ is a closed surface and has the
topology of a torus $\mathbb{T}^2$. A point on $\mathcal{S}$ is specified by
two angles $\vartheta^1$, the angle in the $n_{gs}$-$n_{gd}$ plane, and the
phase $\varphi \equiv \vartheta^2$.  The cylinder's radius $\rho$ is chosen so
as to include only one degeneracy $\mathbf{T_1}$.  It has the same Chern index
as any other surface which includes $\mathbf{T_1}$ :
$c_1^{\band{0}}(\mathcal{S})=q^{\band{0}}(\mathbf{T_1})=+1$ with respect to the
ground level. By deformation, the cylinder offers the advantage to contain the
same helical paths (Eq.\ref{chemins}) as the one on the torus used in
Sec.~\ref{sec:TQCP}.  TQCP can therefore be used exactly in the same fashion
for the topological charge which is quantized as

\begin{align}
\langle
Q\rangle=\frac{2e}{\hbar}\,p\,\hbar\,c_1^{\band{0}}(\mathcal{S})=2e\,p\,,
\end{align}
when averaged over the initial phase $\lambda$ of the helix (defined in
Eq.~\ref{chemins}) making $p$ turns around $T_1$.

\begin{figure*}
\centering
\includegraphics[width=16cm]{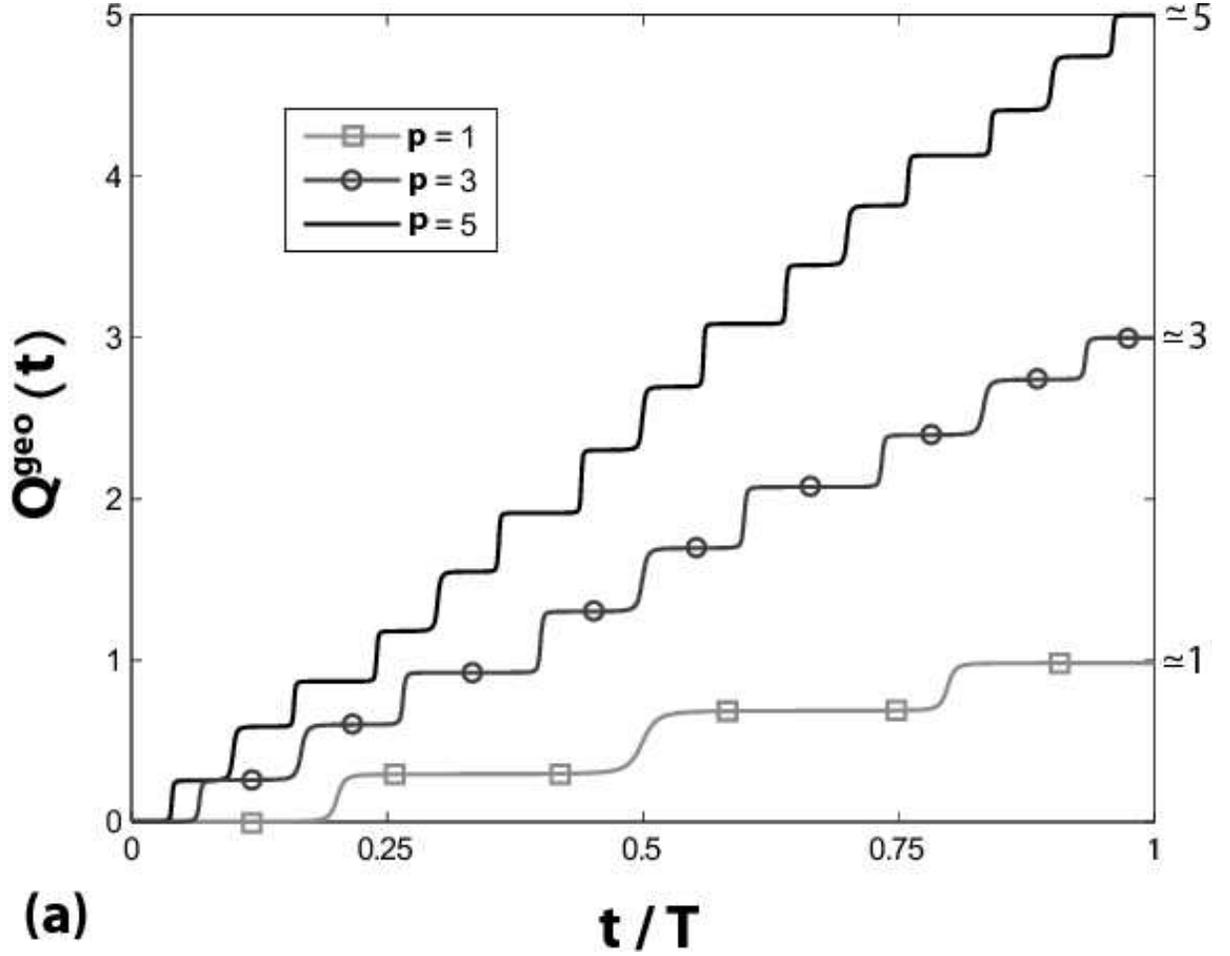}
\caption{(a) Geometric transferred charge $Q^{\rm geo}$ as a
function of $t$ for the helical paths defined in
Eq.~(\ref{chemins}) and $p=1,3,5$. (b) Each time the path crosses
one of the three saddle lines (here plotted for $\varphi=\pi$),
approximatively a third of one charge unit is geometrically
transferred. The computation was done using  a radius $\rho=0.3$,
$\lambda=0$, and the Josephson couplings and charging energies of
a real device: $E_J=6\,\mu eV$, $E_{J0}=8.6\,\mu eV$,
$C_J=0.78\,fF$ and $C_0=0.69\,fF$.} \label{escalier}
\end{figure*}

Since Eq.~\ref{Q_instantaneous} gives the incremental charge transferred, this
quantity can be monitored continuously as one moves along the helical open path
using Eq.~\ref{Q_open}. The Josephson couplings $E_J$ and $E_{J0}\simeq E_J$
lift the degeneracies between the charged states along the three boundary lines
of the hexagons intersecting at $T_1$, which become saddle lines (see
Fig.~\ref{honeycomb} and Fig.~\ref{escalier}-b for the saddle lines of the
ground state). The splitting between the ground state $\k{-}$ and the first
excited state $\k{+}$ is smallest along the saddle lines and are of the order
of $E_J$. As one moves on the helical path, each saddle line is crossed once
per turn around $T_1$. Since the accuracy of the CPP hinges on the ability to
move adiabatically in the ground state manifold, Landau Zener-transition
$\k{-}\rightarrow\k{+}$ when crossing a saddle line are a concern. The
transition probability $P_{L-Z}\simeq e^{-\left(\frac{3\pi}{2}\right)^2
\frac{E_J}{E_C}\frac{E_J}{h\nu_\theta}}$ depends on the ratios
$\frac{E_J}{E_C}$ and $\frac{E_J}{h\nu_\theta}$ which cannot be too small. On
the other hand, when the ratio $\frac{E_J}{E_C}$ is too large, the dynamical
contribution to the charge transferred are more difficult to average out and
the accuracy of the device deteriorates. This is the tradeoff when optimizing
the CPP: a large $E_J$ reduces Landau-Zener tunnelling and allows for a higher
frequency of operation but the Josephson current can be most easily  driven to
zero at small $E_J$. To avoid single electron effects, the charging energy
$E_C$ has to be smaller than the superconducting gap (0.2 meV for Aluminum).
This sets the overall energy scale and most of the parameters: typically values
for $E_C\approx 0.1\,$meV, $E_J\approx 0.05\:E_C$ and $\nu_\theta\approx
100\:$MHz offer a good optimization of the CPP.  The parameters of the helical
path are the radius $\rho$ and the number of turns $p$ in a period
$T_\varphi=\frac{2\pi}{\nu_\varphi}$. For the optimal radius
$\rho=\frac{1-\kappa_0}{2}\approx\frac{1}{3}$, the path intersects the line
between $T_1$ and $T_2$ in the middle.  A smaller radius is equivalent to
reducing $E_C$.

Now that the parameters are known, the charge transferred can be
followed as one moves along the helical paths defined in
Eq.~\ref{chemins}. $p$ is here the number of turns around $T_1$ in
a period $T_\varphi$. Integrating the instantaneous transferred
charge (Eq.~\ref{Q_instantaneous}) for the ground state $\k{-}$
yields the time dependence of $Q^{\rm geo}$
\begin{align}
Q^{\rm geo}(t)=\int_0^t\delta Q^{\rm geo}(t'),
\end{align}
which is plotted as a function of time in Fig.~\ref{escalier}
using the parameters of a real device ($\frac{E_J}{E_C}\approx
0.05$) and Eq.~\ref{Q_instantaneous}. For small $\frac{E_J}{E_C}$
ratios, the charge is transferred in three distinct steps,
corresponding to the transfer of a Cooper pair through each
junction which occurs when crossing the three saddle lines. Two
steps are of height $\simeq\kappa_J$ (external junctions) while
one is of height $\simeq\kappa_0$, yielding a total transferred
charge $\simeq 2\kappa_J+\kappa_0=1$ per turn (in units of $2e$),
as illustrated in Fig.~\ref{escalier} where this quantity is
plotted for different number of turns ($p$). The steps rounding
become more pronounced as $E_J$ increases, and their size more
sensitive to the initial phase value $\varphi(0)=\lambda$.

\begin{center}
\begin{figure*}
\includegraphics[width=0.9\columnwidth]{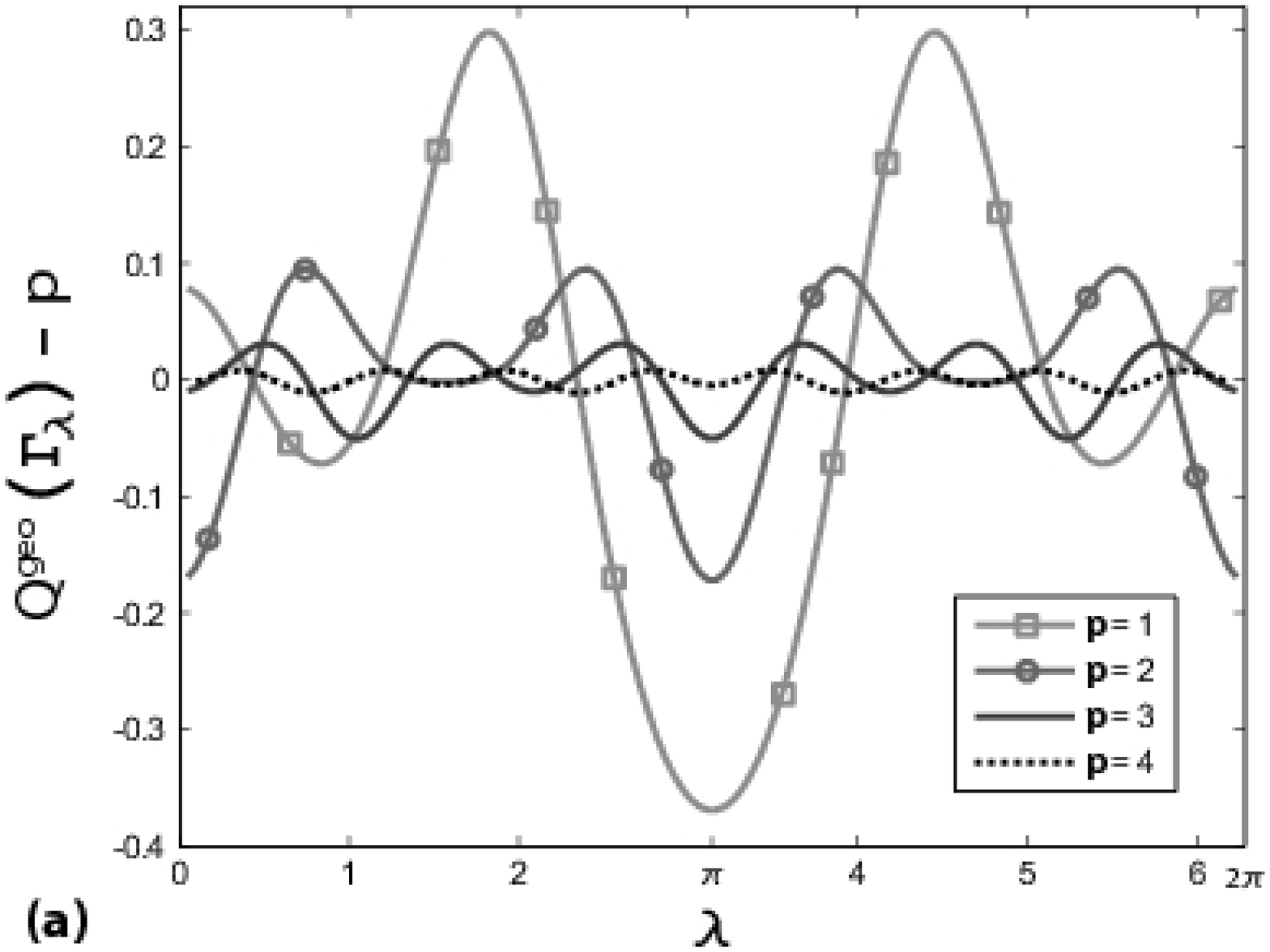}\hspace{1cm}
\includegraphics[width=0.9\columnwidth]{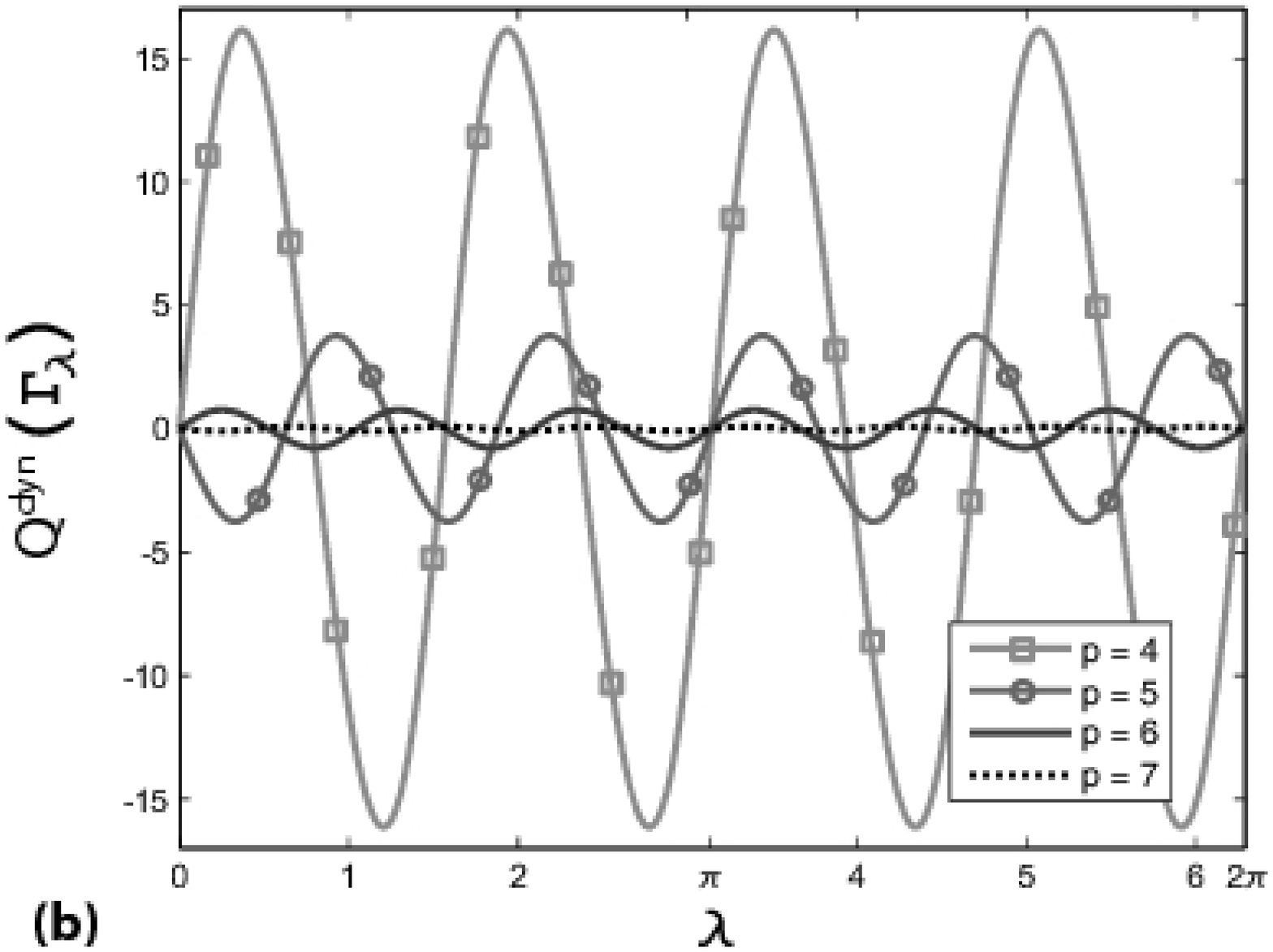}
\caption{(a) Plot of the geometric charge $Q^{\rm geo}$ as a
function of of $\lambda$ for helices with different numbers (1-4)
of windings. (b) Same plot for the dynamical charge $Q^{\rm dyn}$
as a function of $\lambda$ using $4$ to $7$ windings. Parameters:
$\rho=0.3$, $E_J=E_{J0}=60\,\mu$eV and $E_C=114\,\mu$eV.}
\label{lesdeux}
\end{figure*}
\end{center}

Clearly, the charge transferred value $\simeq 2e$ is due to the presence of the
degeneracy with topological charge $+1$. We now verify that quantization
accuracy improves as $p$ increases, a fundamental feature of TQCP:
\begin{align}
Q(\Gamma_\lambda)\underset{p\to\infty}{\longrightarrow}2e\,p\,.
\end{align}
The geometric and dynamical charge transferred in the CPP ground
state can be followed as a function of the initial phase
$\lambda$.  For optimal values of the parameters ($E_C \approx
100\mu$eV and $E_J \approx 3\mu$eV), the errors computed are
small. In the simulation, it is useful to amplify their effects by
choosing the most unfavorable parameters. Using a perturbative
analysis, the dynamical contributions to the pumped charge are of
order $Q^{\rm
dyn}\approx\frac{T_\varphi}{\hbar}\frac{E_J^3}{E_C^2}$ in units of
$2e$ \cite{Pekola99}. Similarly, the deviations of the geometrical
pumped charge from its quantized value scale as
$\frac{E_J^2}{E_C^2}$.  Hence, larger values of $E_J/E_C$ increase
errors.  In Fig.~\ref{lesdeux}, the geometrical and the dynamical
charge transferred are plotted for a ratio $E_J/E_C=0.5$, an order
of magnitude larger than the optimal values. On this figure, the
charges are computed using helical paths around $\mathbf{T_1}$
with different number of windings in a period $T_\varphi$. The
geometrical charge oscillates as a function of $\lambda$ around
the quantized value with an amplitude which decreases rapidly with
the number of windings $p$.  This rate depends mostly on $E_J/E_C$
and on the distance between each windings relative to the distance
to the degeneracy $\mathbf{T_1}$.  This is the main reason to keep
the helix radius close to its optimal value
($\frac{1-\kappa_0}{2}$).

For parameters closer to their optimal value, this decrease can be expressed in
term of $\xi$, the root mean square amplitude of the oscillations. This
quantity is tabulated as in the table \ref{tableau} using $E_J/E_C=0.05$. Above
a few winding the quantization accuracy is very high.  If low frequencies phase
jitters in $\delta\varphi$ are present, the error in the pumped charge will be
of the order of $\xi_{\rm geo}(p) \frac{\delta \varphi}{2\pi}$, which is below
$10^{-8}$ for $p\ge 5$.

\begin{center}
\begin{table}
\begin{ruledtabular}
\begin{tabular}{cccc}
$p$ & $\xi_{\rm geo}$ & $\xi_{\rm dyn}$ & $\xi_{\rm tot}$\\
\hline 1 & $8.7\;10^{-3}$ & $5.4$ &
$5.4$\\ 2 & $2.8\;10^{-3}$ & $4.2\;10^{-3}$ & $5.0\;10^{-3}$\\
3 & $3.8\;10^{-4}$ & $1.3\;10^{-3}$ & $1.3\;10^{-3}$ \\
4 & $3.9\;10^{-5}$ &
$9.1\;10^{-5}$ & $9.9\;10^{-5}$ \\
5 & $4.6\;10^{-6}$ & $5.5\;10^{-6}$ & $7.2\;10^{-6}$\\
\end{tabular}
\end{ruledtabular}
\caption{Mean root squares deviations of the different contributions to the
transferred charge as a function of $p$. Numeric values : $\rho=0.3$,
$\lambda=0$, $E_J=60\,\mu eV$, $E_J/E_C \approx 0.5$ .}\label{tableau}
\end{table}
\end{center}

The average dynamical charge transferred over a period converge also toward
zero when $p$ is sufficiently large, provided the ratio of $\frac{E_J}{E_C}$ is
not too large (say below than $0.05$).  In absence of noise, the periodicity in
$\varphi$ guarantee that it averages out to zero. In the presence of a phase
noise $\delta \varphi$ the cancellation becomes approximate with an error of
order $\xi_{\rm geo}(p)\sim \frac{\delta \varphi}{2\pi}$.

It is not possible to reduce the ratio $E_J/E_C$ arbitrarily to improve
the accuracy, because the gap at the saddle points ($\propto E_J$)
decreases, and the Laudau-Zener tunnelling turn on transition to the
first excited manifold. This introduces the largest source of errors
because the Chern indices of the two lowest eigenvector bundles are
opposite, an issue which is addressed in the concluding section.
\section{\label{sec:adiab-computing}Adiabatic computing}
In the vicinity of a triple point $T_1$, the two lowest states
$\k{\pm(\mathbf{R})}$ form a qubit. In this region of $\mathbb{P}$, the
next level $\k{e(\mathbf{R})}$ lies at an energy of order $\frac{3}{2}
E_J$ above the $\k{\pm}$ doublet, significantly larger than the doublet
splitting. One-qubit operations on this doublet can be implemented
using different schemes.

The simplest one consists in applying microwave pulses to the CPP gates
at the frequency splitting between the $\k{\pm(\mathbf{R})}$ states.
The point $\mathbf{R}$ may be chosen at one of the magical points along
the saddle lines of the hexagons where the system is to first order
insensitive to gate charge and phase fluctuations \cite{Vion02}. But
there are simpler circuits where such operations have been demonstrated
and the CPP is more interesting to implement quantum computation using
adiabatic cycles.  Duan and coworkers\cite{Duan01} have shown how
geometric gates could be implemented on a degenerate two-levels system
using resonant transitions between this $\k{\pm}$ doublet and an
excited level $\k{e}$ provided it is also coupled to an auxiliary level
$\k{a}$. In the CPP, $\k{a}$ could in principle be a higher lying
charge state, but this solution is not as convenient as for atomic
systems.  We prefer to do without the $\k{a}$ state. In this case, the
dynamical contribution associated to the Rabi frequency cannot be
eliminated altogether. As long as this phase shift can be tuned to a
multiple of $\pi$, a geometric gate can be implemented.

The CPP is biased at degeneracy ($\mathbf{T_1}$), and microwave
voltages tuned at the $\k{e}-\k{\pm}$ frequency splitting  are applied
to {\em both} gates. When going to a frame rotating at the same
frequency, the coupling to the $\k{e}$ state becomes time independent
\begin{equation}
\hat{\mathcal{H}}_{\rm mw}(t)=\hbar \Omega \left( \cos
\frac{\zeta}{2}\k{e}\b{-}+ \sin \frac{\zeta}{2} e^{i\chi}
\k{e}\b{+} +{\rm h.c.}\right),
\end{equation}
where $\Omega$ is the main Rabi frequency, and the angle $\zeta$
controls the relative strength of the couplings of the $\k{+}$ and
$\k{-}$ to the $\k{e}$ state, while $\chi$ is their relative phase.
These three quantities can be adjusted by tuning the amplitude and the
phase of the microwave voltages applied to the CPP gates. Here, $2\pi$
phase cycles in $\chi$ are sufficient to generate the gate operations.
In the rotating frame, one of the sate is stationary, while the other
two oscillate at $\pm \Omega$. Adjusting the period of operation $T$ to
eliminate the dynamical phase shift ($e^{i\Omega T}=\pm 1$), a
$2\pi$-cycle in $\chi$ generates the operation

\begin{widetext}
\begin{equation}
G(\zeta,2\pi)=\left(\begin{array}{cc}
\cos(\pi\cos\zeta)\cos\zeta+i\sin(\pi\cos\zeta) &
-\cos(\pi\cos\zeta)\sin\zeta \\
\cos(\pi\cos\zeta)\sin\zeta &
\cos(\pi\cos\zeta)\cos\zeta-i\sin(\pi\cos\zeta)
\end{array} \right)
\end{equation}
\end{widetext}
which covers all the one-qubit gate operation.

Two-qubit geometrical gates can also be considered by coupling two
Cooper pair pumps together.  One way this may be achieved is to close
the two pumps on the same inductance $L$: the current pumped in both
device add up in $L$, shifting the phase $\varphi$ by an amount of
order $LI_j$ where $I_j$ is the current through the $j$-th CPP.

In practice, adiabatic computing faces a number of difficulties. Even
for one-qubit operations three or more states have to be degenerate.
But any low frequency flux or charge noise moves the area in parameter
space where the degeneracy occurs and the shifts in parameters rotates
the eigenvectors rapidly. While this has little impact for topological
quantization, this deteriorates the performance of geometrical gates.
The accuracy with which parameters are to be controlled is also
considerably higher than for usual quantum gates.  These are some of
the reason why geometrical gates have not yet been demonstrated.

\section{Conclusions}
Most limitations for adiabatic computing are irrelevant to topological
quantization which is robust against most perturbations.  For a CPP, charge or
phase noise is considered slow when most of its spectrum is below the frequency
of operation of the pump $\nu_\theta$ (100 MHz is a realistic number). Such
noise source add a random component to the controlled voltage or the magnetic
flux and the path in $\mathbb{P}$ no longer generates a cylinder, but a more
irregular surface.  As far as the geometrical charge is concerned, this has
almost no effect, as long as the resulting surface still encloses the
topological point $\mathbf{T_1}$.  Similarly if the junction capacitances or
Josephson couplings fluctuate in time, the position of the point $\mathbf{T_1}$
fluctuates in $\mathbb{P}$, and this has little effect as long as this shift is
small compared to the cylinder's radius.  In presence of low frequency noise,
dynamical contributions no longer cancel exactly.  On the other hand, the
errors are random and can be averaged using long integration times.

If charge or phase noise has frequency components at the splitting between the
two lowest states $\k{\pm(\mathbf{R})}$ or if Landau Zener transition occur at
one of the three saddle point crossings (see Sec.~\ref{sec:charge-quantization}
for a discussion), the system will spend a fraction of the time in the first
excited state.  Since the Chern index of the $\k{+}$ state is opposite to the
ground state, this will introduce an error in the pumped charge proportional to
the relative time spent in the excited state. This is why high frequency noise
must be thoroughly filtered and the pumped speed adjusted to quench
Landau-Zener transitions.

This study of quantization by {\em controlled path}, although conducted
around CPP circuits, is quite general. The necessary criteria which
have been stated in the introduction, are now discussed in more detail

\begin{enumerate}
\item The adiabatic condition can only be satisfied for discrete
spectra. Furthermore, adiabaticity only holds well for a quantum state if its
energy splittings with other levels are sufficient.

\item The presence of isolated degeneracies between two lowest
states states is required.  For a complex Hamiltonian, this is possible only if
it depends on three continuous parameters.  In quantum circuits, tunable
parameters are typically gate voltages and magnetic fluxes. For Cooper pair
pumps, the parameters are two gates charges and a phase (CPP) or two phases and
a charge (cf. the sluice pump \cite{Pekola99}). The parameters are used to
generate \textit{controlled paths} in the three-dimensional parameter space
$\mathbb{P}$.  Additional parameters are useful for adiabatic computations
(section V).  There are two recipes to locate the degeneracies.  The first one
hinges on their topological signature on the Berry's phase ($\pi$). As argued
by A.J. Stone \cite{Stone76}, one can divide the parameter space $\mathbb{P}$
in small volumes around which Berry's phase is computed.  If a single
degeneracy exists inside the loop, Berry's phase picks it up. Another method is
to detect the vorticity of the eigenvector bundle in the vicinity of a
degeneracy (see sections \ref{sec:fiber-bundle} and \ref{sec:CPP}):  in this
case the ``mirror symmetry'' and ``T-symmetry'' are broken for the ground
state, i.e. they flip the eigenvector bundles intersecting at the degeneracy
(Kramers symmetry).

\item Only observable proportional to the partial derivative of
the Hamiltonian $\hat{\mathcal{H}}$ with respect to one of the
tunable parameters $\vartheta^2$ ($\varphi$ for the CPP) can be
quantized using TQCP. In this case, its averaged expectation value
over a closed surface $\mathcal{S}$ around a degeneracy is
proportional to the Chern index of this surface. When the variable
is periodic, we may choose this surface with the topology of a
torus. A path sweeping this surface or covering it densely is the
{\em controlled path} expressing the quantization.
\end{enumerate}

Although topological quantization is quite robust, transitions to the
first excited state (induced by fluctuations or Landau-Zener processes)
are problematic because the Chern indices of the two lowest levels have
opposite sign.

The above criteria only concern geometrical contributions. But the
dynamical evolution of the observable may also contribute.  An accurate
quantization is possible if they can be eliminated through symmetries
or other schemes. This depends to some extend of the physical problem
on hand.  For quantum circuits, the following condition is sufficient
to average them out:

\begin{enumerate}
\setcounter{enumi}{3} \item The Hamiltonian and the surface $\mathcal{S}$ are
periodic in the parameter $\varphi$, or have some symmetry with respect to
$\varphi$, such that the dynamical contribution averages to zero.
\end{enumerate}

The value of the topological quantization is its strong robustness to
adiabatic parameters fluctuations, the Chern index of the surface
enclosing the degeneracy being the quantum number of the quantity of
interest.  For a Cooper pump, the magnitude of the current generated
\begin{equation}
I= 2e\, c_1^{\band{0}} (\mathbf{T_i})\, \nu_\theta
\end{equation}
is of order 30 pA for realistic values of the parameters.

One of the potential application for a Cooper pair pump is as an accurate
current source\cite{Vartiainen07} to close the metrology-triangle
\cite{Piquemal00}, relating frequency to voltage through the Josephson effect,
voltage to current through the quantum Hall effect and frequency to current
through topological pumping. It is interesting to note that all these effects
are the result of topological quantization \cite{Thouless83, Thouless82,
Kohmoto84, Goryo07}.  In order to be useful in this context, where the pumped
current feeds a Hall bar, a higher current is needed (100 nA or
higher\cite{Delahaye03}). Large gain amplification schemes are being designed
to fulfill this condition. Another method for a current-frequency conversion
relies on Bloch oscillation\cite{Averin85}. This method has been recently
demonstrated by the quantronium group \cite{Nguyen07}. This method, also based
on a topological quantization (Thouless pumping\cite{Thouless83}), also appears
to be quite promising.

To conclude, we have stressed the importance of topological quantization in
superconducting circuits. We showed how closed paths can be chosen in parameter
space to generate a surface on which the ``charge'' $Q$ is quantized by a Chern
index. We have also shown how the charge transferred on any path could be
obtained, giving also a local picture which cannot be derived from the usual
global geometrical picture.  Finally, the technique developed around
superconducting circuits is quite general and can be applied to any problems in
which the criteria listed in the introduction and conclusion are satisfied.

\begin{acknowledgments}
R. Leone is supported by a fellowship of the Rh\^{o}ne-Alpes region (Micro-Nano
cluster) We are quite grateful to Fr\'ed\'eric Faure for many enlightening
discussions.
\end{acknowledgments}

\end{document}